\documentclass[useAMS,usenatbib]{mnras}
\usepackage{natbib}
\usepackage{epsfig}
\usepackage{scrtime}
\usepackage{graphicx}   % Including figure files
\usepackage{amsmath}    % Advanced maths commands
\usepackage{amssymb}    % Extra maths symbols
\usepackage{float}
\usepackage[caption = false]{subfig}
\usepackage{multirow}
\usepackage{rotating}
\usepackage{ulem}
\usepackage{booktabs}
\usepackage{wasysym}
\usepackage{xcolor}

\usepackage{soul}

\newcommand{\mincir}{\raise-3.truept\hbox{\rlap{\hbox{$\sim$}}\raise4.truept\hbox{$<$}\ }}

\usepackage{hyperref}
\usepackage{tikz}

\definecolor{lime}{HTML}{A6CE39}
\DeclareRobustCommand{\orcidicon}{
	\begin{tikzpicture}
	\draw[lime, fill=lime] (0,0) 
	circle [radius=0.13] 
	node[white] {{\fontfamily{qag}\selectfont \tiny ID}};
	\draw[white, fill=white] (-0.0625,0.095) 
	circle [radius=0.007];
	\end{tikzpicture}
	\hspace{-2mm}
}

\foreach \x in {A, ..., Z}{\expandafter\xdef\csname orcid\x\endcsname{\noexpand\href{https://orcid.org/\csname orcidauthor\x\endcsname}
			{\noexpand\orcidicon}}
}
\usepackage[T1]{fontenc}
\usepackage{ae,aecompl}

\usepackage{newtxtext,newtxmath}
\title[Planetary material around WDs]{Synthetic X-ray emission from white dwarf accreting planetary material}

\author[S.~Estrada-Dorado et al.]{S. Estrada-Dorado\thanks{E-mail: s.estrada@irya.unam.mx}$^{1\orcidD}$, V.~Lora$^{1,2\orcidB}$, J.~A.\,Toal\'{a}$^{1\orcidA}$\thanks{Visiting astronomer at the Instituto de Astrof\'{i}sica de Andaluc\'{i}a (IAA-CSIS,
Spain) as part of the Centro de Excelencia Severo Ochoa 2022 Visiting-Incoming
programme.}, 
A.~Esquivel$^{2\orcidC}$, 
M.~A.~Guerrero$^{3\orcidE}$, 
R.~F.~Maldonado$^{1}\orcidG$ 
\newauthor and 
Y.-H.\,Chu$^{4\orcidF}$\\
$^{1}$Instituto de Radioastronom\'{i}a y Astrof\'{i}sica, UNAM, Antigua Carretera a P\'{a}tzcuaro 8701, Ex-Hda. San Jos\'{e} de la Huerta, Morelia 58089, Mich., M\'{e}xico\\
$^{2}$Instituto de Ciencias Nucleares, UNAM, Apartado postal 73-543, 04510 Ciudad de M\'{e}xico, Mexico\\
$^{3}$Instituto de Astrof\'{i}sica de Andaluc\'{i}a, CSIC, Glorieta de la Astronom\'{i}a S/N, Granada E-18008, Spain\\
$^{4}$Institute of Astronomy and Astrophysics, Academia Sinica, No.\ 1, Section 4, Roosevelt Road, Taipei 10617, Taiwan 
}

\begin{document}
%date{v0 -- Compiled at \thistime\ hrs  on \today\ }
%\pagerange{\pageref{firstpage}--\pageref{lastpage}} \pubyear{2020}
\maketitle
\label{firstpage}

\begin{abstract}
\noindent
The emission of hard X-rays associated with white dwarfs (WD) can be generated by the presence of a stellar companion either by the companion's coronal emission or by an accretion disk formed by material stripped from the companion. Recent studies have suggested that a Jupiter-like planet can also be donor of material whose accretion onto the WD can generate hard X-rays. We use the {\sc guacho} code to reproduce the conditions of this WD-planet scenario. With the example of the hard X-ray WD KPD\,0005+5106, we explore different terminal wind velocities and mass-loss rates of a donor planet for a future network of simulations to investigate the luminosity and the spectral and temporal properties of the hard X-ray emission in WD-planet systems. 
Our simulations show that the material stripped from the planet forms a disk and accretes onto the WD to reach temperatures high enough to generate hard X-rays as usually seen in X-ray binaries with low-mass companions. For high terminal wind velocities, the planet material does not form a disk, but it rather accretes directly onto the WD surface. The simulations reproduce the X-ray luminosity of another X-ray accreting WD (G\,29$-$38), and only for some times reaches the hard X-ray luminosity of KPD\,0005+5106. The X-ray variability is stochastic and does not reproduce the period of KPD\,0005+5106, suggesting that additional physical processes (e.g., hot spots resulting from magnetic channelling of the accreting material) need to be explored. 
\end{abstract}


\section{Introduction}
\label{sec:intro}

White dwarfs (WDs) represent the final stage of the evolution of low- and intermediate-mass stars (1~M$_\odot \lesssim M_\mathrm{i} \lesssim 8$~M$_\odot$). 
These stars lose most of their initial mass through a dense and slow wind \citep[$\dot{M}\lesssim$10$^{-4}$~M$_\odot$~yr$^{-1}$, $v \approx10$~km~s$^{-1}$; see][and references therein]{Scicluna2022} when evolving through the asymptotic giant branch (AGB) phase. Having exposed their cores, they become hot enough to develop fast winds \citep[$v_\infty \gtrsim$10$^{3}$~km~s$^{-1}$;][]{Guerrero2013} and strong UV photon fluxes. The combinations of these effects create what we know as planetary nebulae \citep[PNe;][]{Kwok2000} whose final fate is to expand and mix into the interstellar medium (ISM) whilst the progenitor star evolves along the WD cooling track.

As stars evolve, their planetary systems co-evolve. In particular the extreme variation of the stellar and wind parameters during the evolution of Solar-like stars will create harsh environments that may endanger the survival of planetary systems. 
It has been proposed that planets located $\lesssim$1~AU from their host stars (e.g., Mercury or Venus) will be evaporated during the main sequence phase \citep{Rao2021} or engulfed during the inflation of the stellar outer layers during the AGB phase \citep{Privitera2016}.  
Those located farther away will spiral in or migrate out as a result of the reduction of the gravitational potential of the star. In addition, any remaining planet could be photoevaporated by the UV flux during the post-AGB phase \citep{Villaver2007}. Against all odds, planets and planetary material have been discovered orbiting WD stars \citep[see, e.g.,][]{Veras2021}.

For decades, the IR excess observed from WDs \citep[see, e.g.,][]{Zuckerman1987,Hansen2006,Bilikova2012,Xu2020} 
and the presence of metals in the atmospheres of cool degenerate DZ WDs \citep[see, e.g.,][]{Jura2003,Jura2008,Farihi_etal2010} have been suggested to be a signature of the presence of planets or the remains of planetary systems. 
After the discovery of a Jupiter-like planet around WD\,1856+534 \citep{Vanderburg2020,Alonso2021}, several works have addressed the search and discovery of planets, planetary material and other sub-stellar objects around WDs \citep[see, e.g.,][]{Blackman2021,Brandner2021,Fitzmaurice2023,Martin2021,VanGrootel2021,vanRoestel2021, Kosakowski2022,Walters2022}.

Recently, X-ray observations have been used to suggest the presence of planets or planetary material around WDs.
\citet{Chu2021} presented the analysis of X-ray emission associated with putative single hot ($T_\mathrm{eff}>10^{5}$~K) WDs. 
They used {\it Chandra} and {\it XMM-Newton} observations of KPD\,0005+5106 to unveil variability in the hard X-ray band (0.6--3.0 keV) with a period of 4.7$\pm$0.3~hr in the WD. In general, hard X-ray emission from WDs is produced by either a companion's coronal activity or the acretion of a companion's material onto a WD \citep[see][]{ODwyer2003,Chu2004,Bilikova2010}. However, using H$\alpha$ and multi-instrument IR observations of KPD\,0005+5106, \citet{Chu2021} were able to reject any stellar companion as late as M\,8V. 
Assuming that the periodicity of the hard X-ray emission detected from KPD\,0005+5106 is the orbital period of a sub-stellar companion (a M9V star, a T-type brown dwarf and a Jupiter-like planet), \citet{Chu2021} estimated that only a Jupiter-like planet would be able to fill its Roche Lobe to transfer material to the WD. Using the hard X-ray luminosity of KDP\,0005+5106 ($L_\mathrm{X,hard}=3\times10^{30}$~erg~s$^{-1}$) they estimated an instantaneous accretion rate of $\dot{M}_\mathrm{ac}$=$1.45\times10^{14}$~g~s$^{-1}$(=$2.3\times10^{-12}$~M$_\odot$~yr$^{-1}$).

More recently \citet{Cunningham2022} used {\it Chandra} observations of the WD G\,29–38 to report the detection of X-ray emission. 
G\,29-38 is a widely studied metal-polluted WD of spectral type DAZ \citep{KPS1997}, which has been known to exhibit IR excess \citep{Zuckerman1987,Graham1990} that has been attributed to the accretion from a dust-rich disk \citep{Jura2003}.
\citet{Cunningham2022} suggested that the X-ray emission is produced by accretion of the planetary material onto G\,29-38. 
Our recent analysis of archival {\it XMM-Newton} observations of G\,29-38 \citep{EstradaDorado2023} confirmed the X-ray emission from G\,29-38, with an X-ray luminosity in the 0.3--7.0~keV energy band of $L_\mathrm{X}=(8.3\pm4.1)\times10^{25}$~erg~s$^{-1}$,
which implies an accretion rate of $\dot{M}_\mathrm{ac}=4.01\times10^{9}$~g~s$^{-1}$(=6.45$\times10^{-17}$~M$_\odot$~yr$^{-1}$).

In this paper we present a set of radiation-hydrodynamic simulations of mass-losing planets orbiting a WD to study the production of X-ray emission. 
The physical properties of the circumstellar material derived from our simulations are post-processed using the {\sc Chianti} database \citep{DelZanna2021} to derive X-ray properties. 
We discuss our results in the light of previous observations of putatively single WDs with X-ray emission such as the metal-polluted WD G\,29$-$38 and the hard X-ray-emitting WD KPD\,0005+5106.

This paper is organized as follows. In Section~\ref{sec:simulation} we describe the code and initial conditions of the simulations. In Section~\ref{sec:results} we present our numerical results and the creation of synthetic X-ray properties and our discussion is presented in Section~\ref{sec:discussion}. Finally, a summary of the work is presented in Section~\ref{sec:summary}.

\section{Numerical simulations}
\label{sec:simulation}

\subsection{The {\sc guacho} code}

We use the well-tested radiation-hydrodinamic 3D code {\sc guacho} \citep{Esquivel2009,Esquivel2013} to model the interaction of a mass-losing Jupiter-like planet around a WD. 
{\sc guacho} includes a modified version of the ionizing radiation transfer presented by \citet{Raga2009}. 
It solves the gas-dynamic equations with a second order accurate Godunov-type method using a linear slope-limited reconstruction and the Harten-Lax-van Leer-Contact (HLLC) approximate Riemann solver \citep{Toro1994} implemented on a uniform Cartesian grid.

Simultaneously with the Euler equations, the code solves the rate equation for neutral and ionized hydrogen
\begin{equation}
    \frac{\partial n_\mathrm{HI}}{\partial t} + \nabla \cdot (n_\mathrm{HI}\textbf{u}) = n_\mathrm{e}n_\mathrm{HII}\alpha(T)- n_\mathrm{HI}n_\mathrm{HII} C(T)-n_\mathrm{HI}\phi, 
\end{equation}
where \textbf{u} is the flow velocity, 
$n_\mathrm{e}$, $n_\mathrm{HI}$ and $n_\mathrm{HII}$ are the electron, neutral hydrogen and ionized hydrogen number densities, respectively, 
$\alpha(T)$ is the recombination coefficient, 
$C(T)$ is the collisional ionization coefficient, and 
$\phi$ is the H photoionization rate due to a central source. 
The photoionizing rate is computed with the Monte-Carlo ray tracing method described by \citet{Esquivel2013} and \citet{Schneiter2016}.

The ionization fraction is defined as:
\begin{equation}
    \chi = \frac{n_\mathrm{HII}}{n_\mathrm{HI} + n_\mathrm{HII}},
\end{equation}
with the total density defined as $n_\mathrm{HI} + n_\mathrm{HII}$. The ionization fraction is used to estimate the radiative cooling, which is added to the energy equation using the prescription described by \citet{Esquivel2013}.
In this exploratory work, magnetic fields are not included, yet its effects can be important.  
These will be reported in a subsequent paper.

\subsection{Initial conditions}
\label{sec:initial_conditions}

All simulations presented here are run in a 3D $600\times600\times600$ cells cartesian grid on a box with physical size $0.015\times0.015\times0.015$~AU$^3$, i.e.\ the simulations have cell resolution of 2.5$\times$10$^{-5}$ AU ($\approx5\times10^{-3}$ R$_\odot$).
We start the simulations adopting a uniform medium with number density of $n_0=1$~cm$^{-3}$ 
and a temperature of $T_0=10^4$~K. 

We simulate a mass-losing planet with initial mass of 1 $\mathrm{M_{\jupiter}}$ orbiting a WD in a circular orbit. The WD is placed at the centre of the numerical grid and the orbital period is set to 4.7~hr as found by the variability of the hard X-ray emission reported in \citet{Chu2021}. This corresponds to
an orbital separation of 5$\times$10$^{-3}$~AU ($\approx$1.07~R$_{\odot}$). The WD properties are set to values  reported for KPD\,0005+5106: effective temperature of $T_\mathrm{eff} = 200,000$~K, no stellar wind, and mass of 0.6~M$_\odot$ \citep{Werner2015}. \\

We ran different simulations varying the mass-loss rate ($\dot{M}$) and outflow velocity ($v_{\infty}$) of the orbiting Jupiter-like planet. Run A adopts a mass-loss rate of $\dot{M}$=2.30$\times$10$^{-12}$~M$_\odot$~yr$^{-1}$, which corresponds to the accretion value estimated in \citet{Chu2021}, and a mass outflow velocity of $v_{\infty}$=60~km~s$^{-1}$ (the escape velocity for this planet). With the goal to explore the effects on the luminosity and accretion rate onto the WD, other runs are performed varying these two values as listed in Table~\ref{tab:simulations}. In all cases, the adopted mass loss for the Jupiter-like planet is isotropic.

\begin{table}
\begin{center}
\caption{Wind parameters (velocity and mass-loss rate) of the mass-losing Jupiter-like planet orbiting a WD. The parameters of the WD remain unchanged in all the simulations. See Sec.~\ref{sec:initial_conditions} for details.}
\begin{tabular}{crcc}
\hline
Run & \multicolumn{1}{c}{$v_{\infty}$}  & \multicolumn{2}{c}{$\dot{M}$} \\
\cmidrule(lr){3-4}
    & \multicolumn{1}{c}{(km~s$^{-1}$)} & (M$_\odot$~yr$^{-1}$) & (g~s$^{-1}$)\\
\hline
A &  60~~~~ & 2.30$\times$10$^{-12}$ & 1.45$\times10^{14}$\\
B &  60~~~~ & 2.30$\times$10$^{-11}$ & 1.45$\times10^{15}$\\
C &  60~~~~ & 2.30$\times$10$^{-10}$ & 1.45$\times10^{16}$\\
D & 600~~~~ & 2.30$\times$10$^{-12}$ & 1.45$\times10^{14}$ \\
\hline
\end{tabular}
\label{tab:simulations}
\end{center}
\end{table}

\section{Results}
\label{sec:results}

\begin{figure*}
\begin{center}
\includegraphics[angle=0,width=\linewidth]{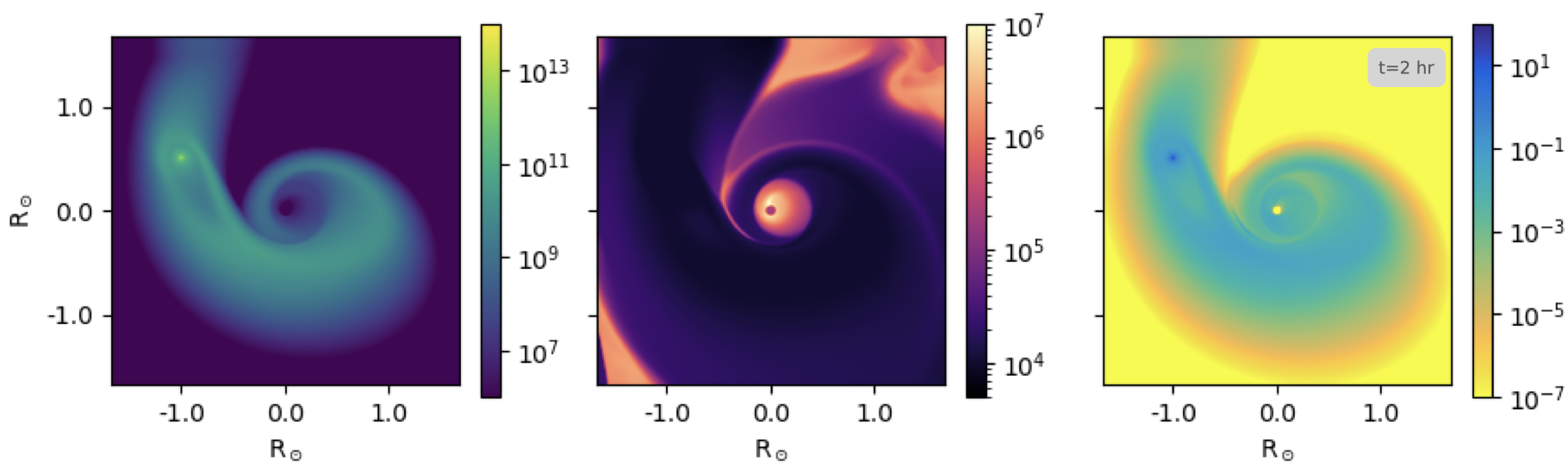}\\
\includegraphics[angle=0,width=\linewidth]{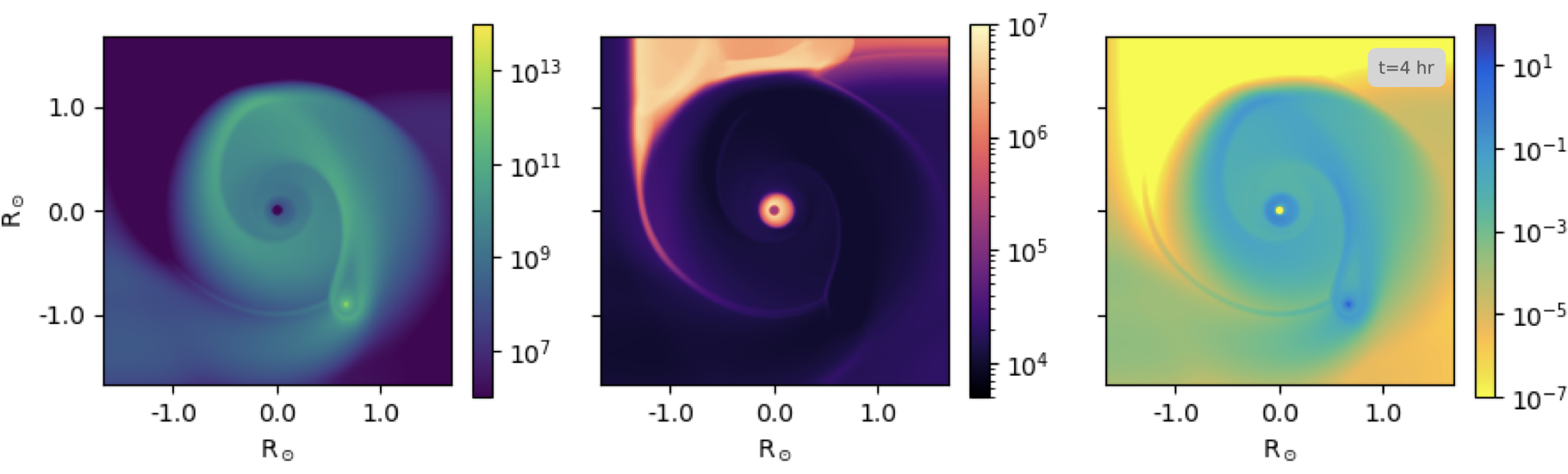}\\
\includegraphics[angle=0,width=\linewidth]{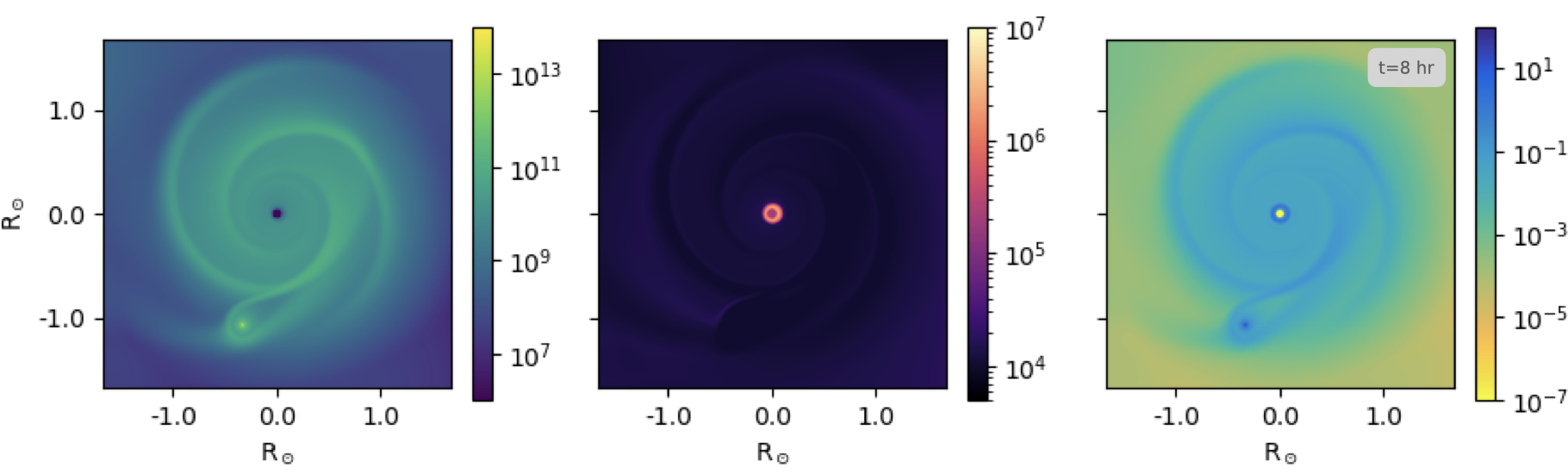}\\
\includegraphics[angle=0,width=\linewidth]{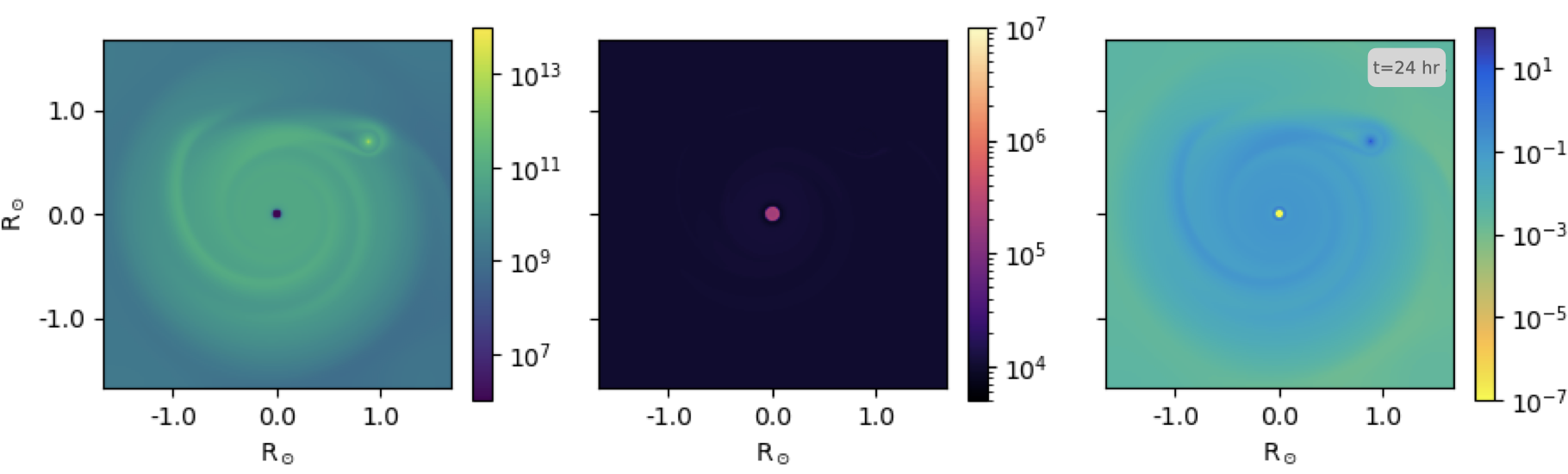}
\caption{Face-on view of the middle plane of the number density $n$ (cm$^{-3}$ - left), temperature $T$ (K - middle) and pressure (erg~cm$^{-3}$ - right) of Run A. 
From top to bottom we show the results after 2, 4, 8 and 24 hours of evolution, respectively.}
\label{fig:A_simfront}
\end{center}
\end{figure*}

\begin{figure*}
\begin{center}
\includegraphics[angle=0,width=\linewidth]{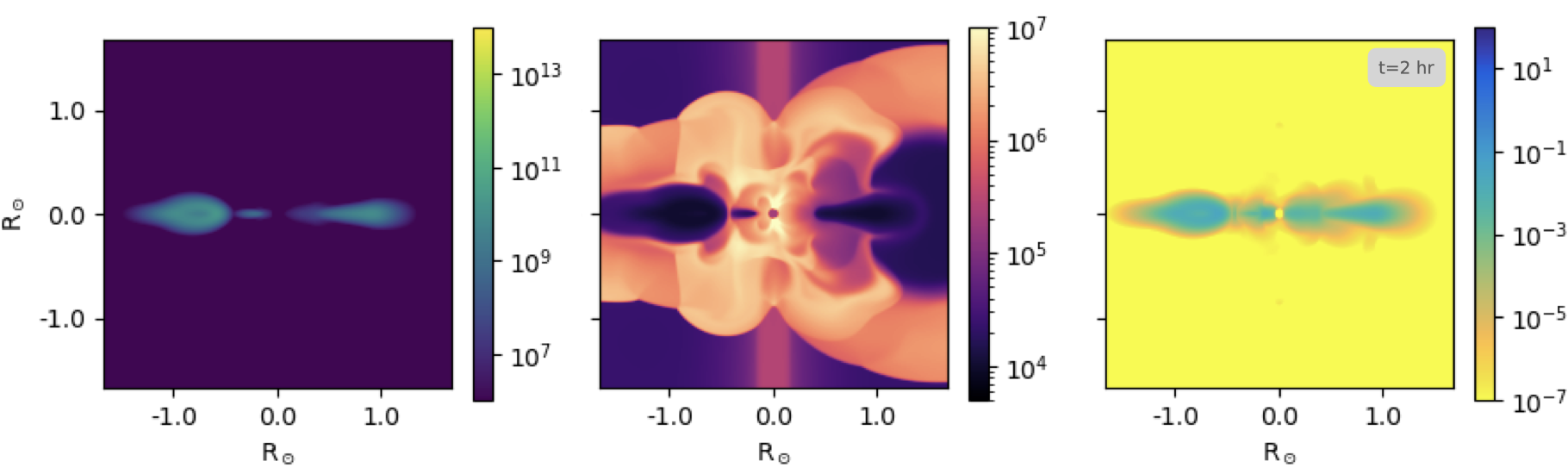}\\
\includegraphics[angle=0,width=\linewidth]{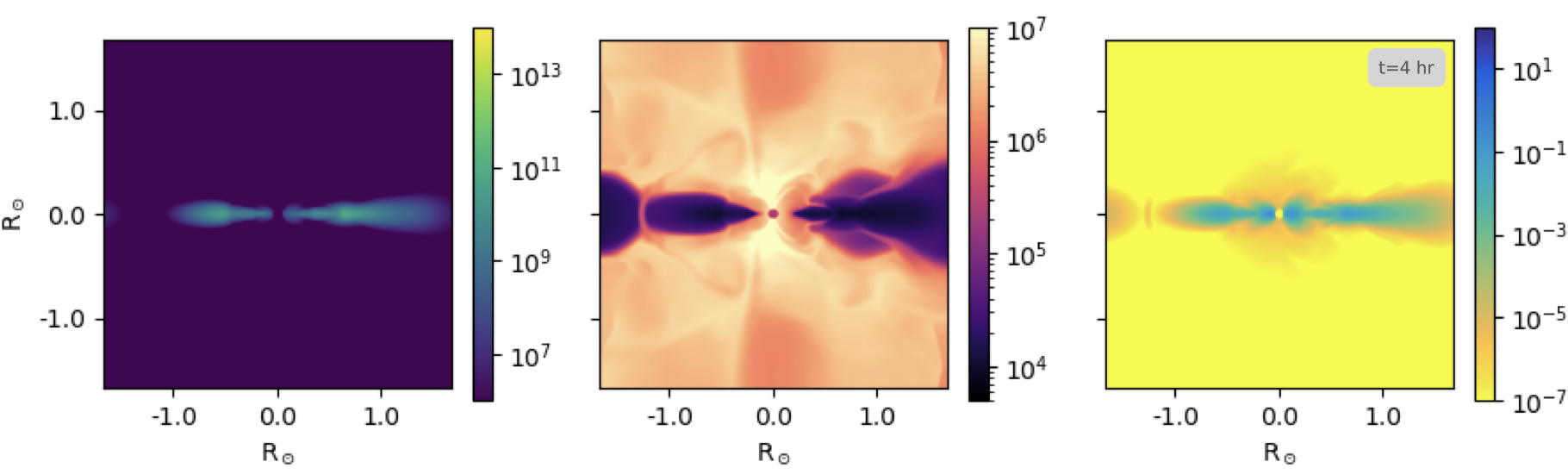}\\
\includegraphics[angle=0,width=\linewidth]{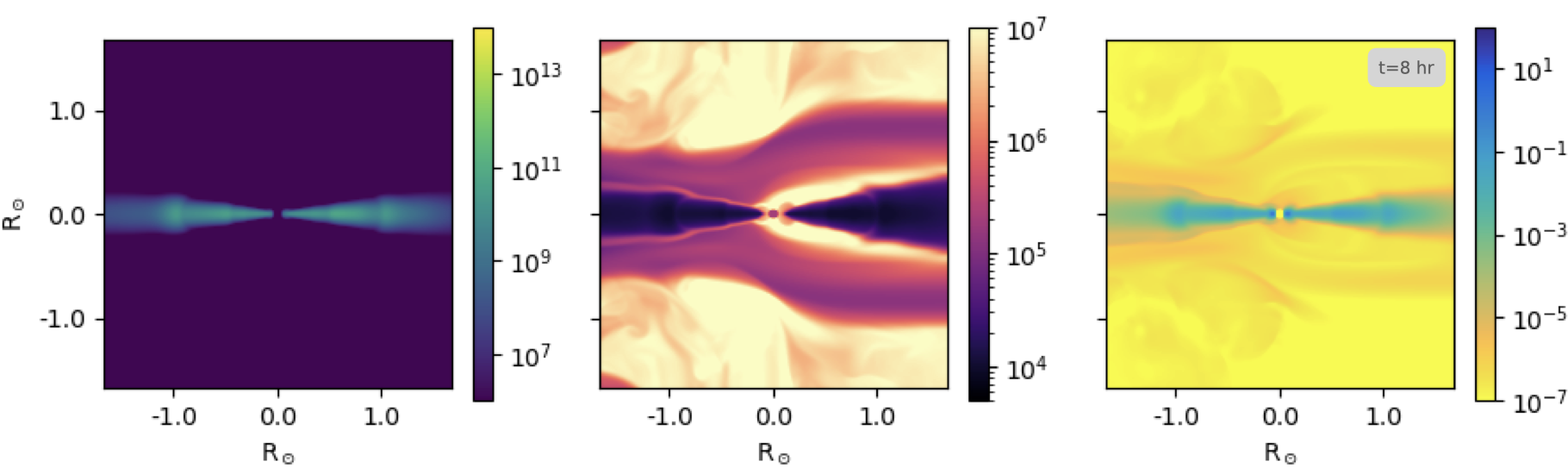}\\
\includegraphics[angle=0,width=\linewidth]{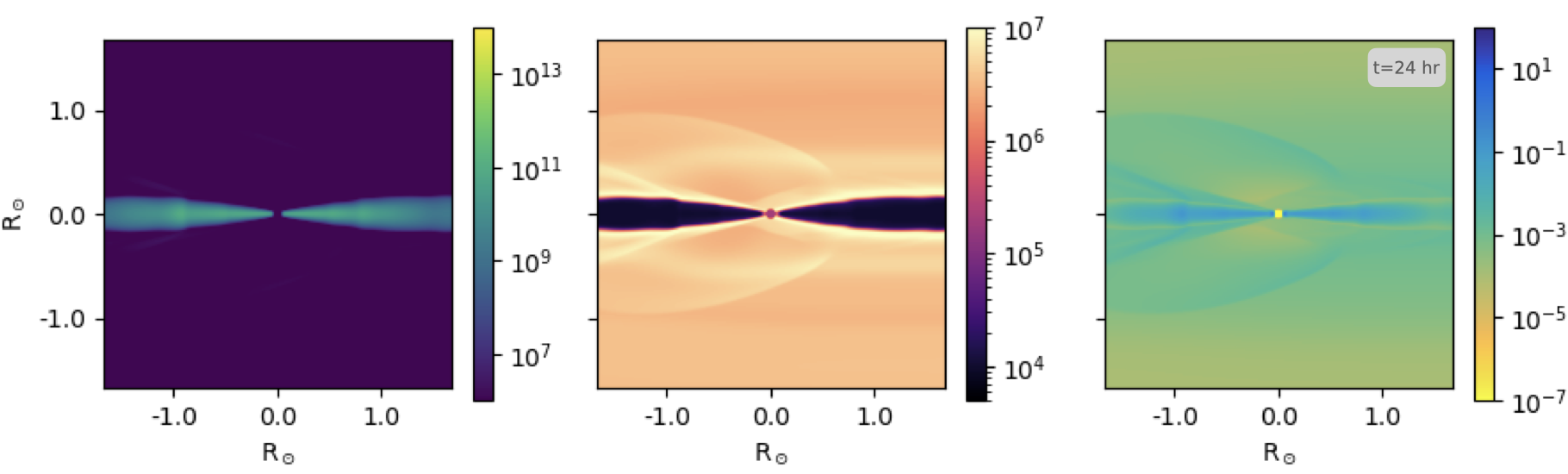}
\caption{Same as Figure~\ref{fig:A_simfront} for the edge-on view of the middle plane.}
\label{fig:A_simside}
\end{center}
\end{figure*}

\begin{figure*}
\begin{center}
\includegraphics[angle=0,width=0.8\linewidth]{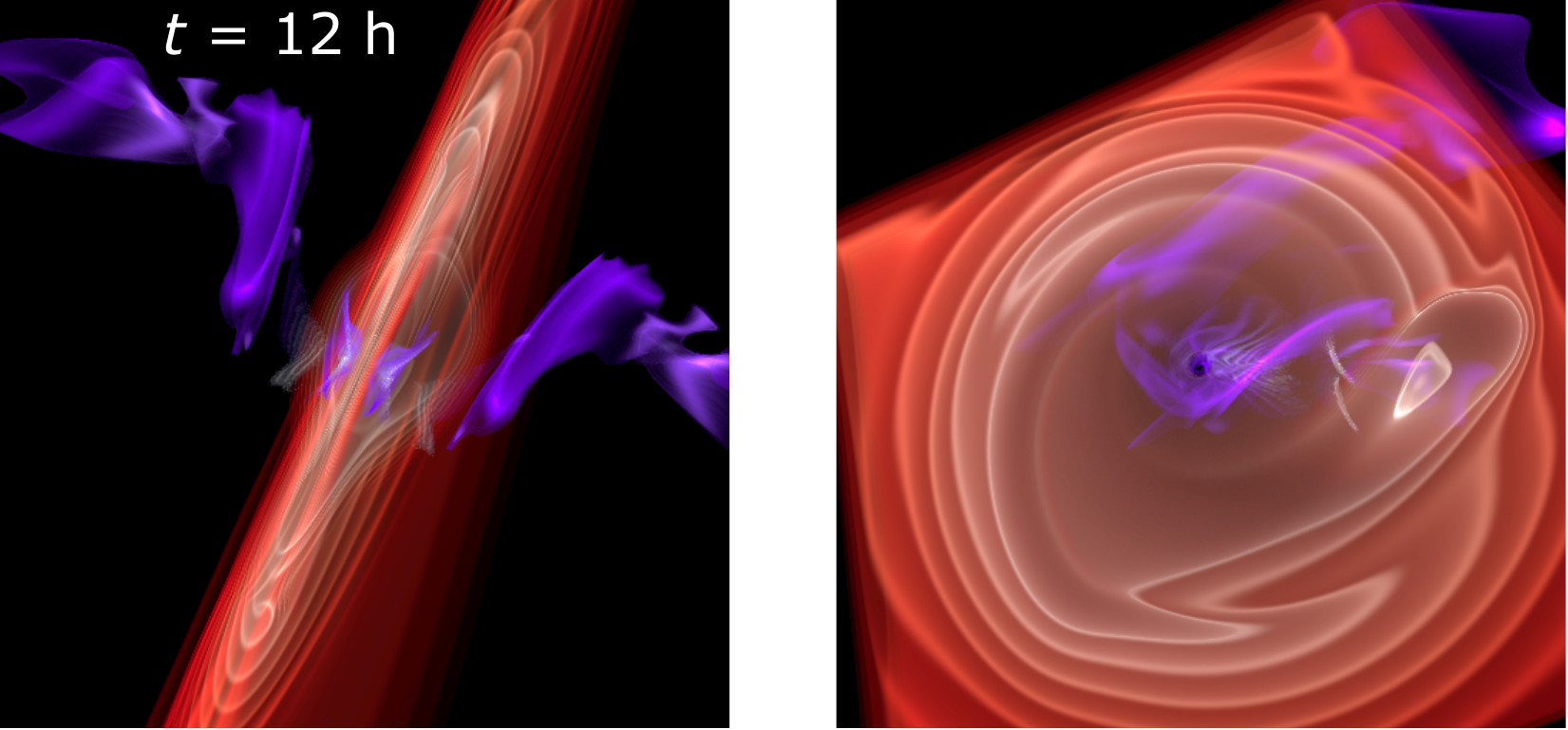}
\caption{Visualization of rendered image of the accretion disk obtained from Run C after 12~h of evolution. The red shade represents material with number density in the 10$^{9}$--10$^{12}$~cm$^{-3}$ range, whilst the purple strings represent material hot enough to produce hard X-rays ($T>10^{7}$~K). A video illustrating the formation and evolution of the disk is presented as supplementary material to this article.}
\label{fig:rendered}
\end{center}
\end{figure*}

The results from Run A are shown in Figures~\ref{fig:A_simfront} and \ref{fig:A_simside}, which correspond respectively to the face-on and edge-on views of the number density $n$, temperature $T$ and pressure $p$. Note that the figures present cuts at the central plane and do not correspond to integrated images along the line of sight.

Figures~\ref{fig:A_simfront} and \ref{fig:A_simside} illustrate the early formation of an accretion disk around the WD. 
Material is lost by the orbiting Jupiter-like planet isotropically and is gravitationally-trapped by the WD. Fig.~\ref{fig:A_simfront} shows that the gas does not fall directly into the WD, but spirals-in towards it creating spiral patterns of shocked gas. 
After 10 hours of evolution the disk around the WD reaches a steady state, with no further significant morphological changes.

Fig.~\ref{fig:A_simside} shows that once the disk is fully formed, its mid-plane region is composed by ionized material with temperatures of 10$^{4}$~K surrounded by a thin layer of slightly higher temperatures ($\sim10^{5}$~K). 
Material in the innermost regions of the disk can be extremely hot shocked up to X-ray-emitting temperatures of $\lesssim10^{7}$~K \citep[similar to the boundary layer of the accretion disk of dwarf novae and non-magnetic cataclysmic variables,][]{PR1985}. 

As illustration, Fig.~\ref{fig:rendered} presents a 3D renderization of Run C after 12~h of the simulation, where the red shade represents material with density large enough, in the 10$^9$--10$^{12}$~cm$^{-3}$ range, to create an accretion disk. On the other hand, the purple shade strings show material with temperature high enough to produce hard X-rays ($T>10^7$~K). This figure shows that X-ray-emitting gas is leaving the accretion disk but from the innermost regions, close to the WD. The X-ray-emitting material leaves the accretion disk in a bipolar stream of gas, but this is not collimated.

\begin{figure*}
\begin{center}
\includegraphics[angle=0,width=\linewidth]{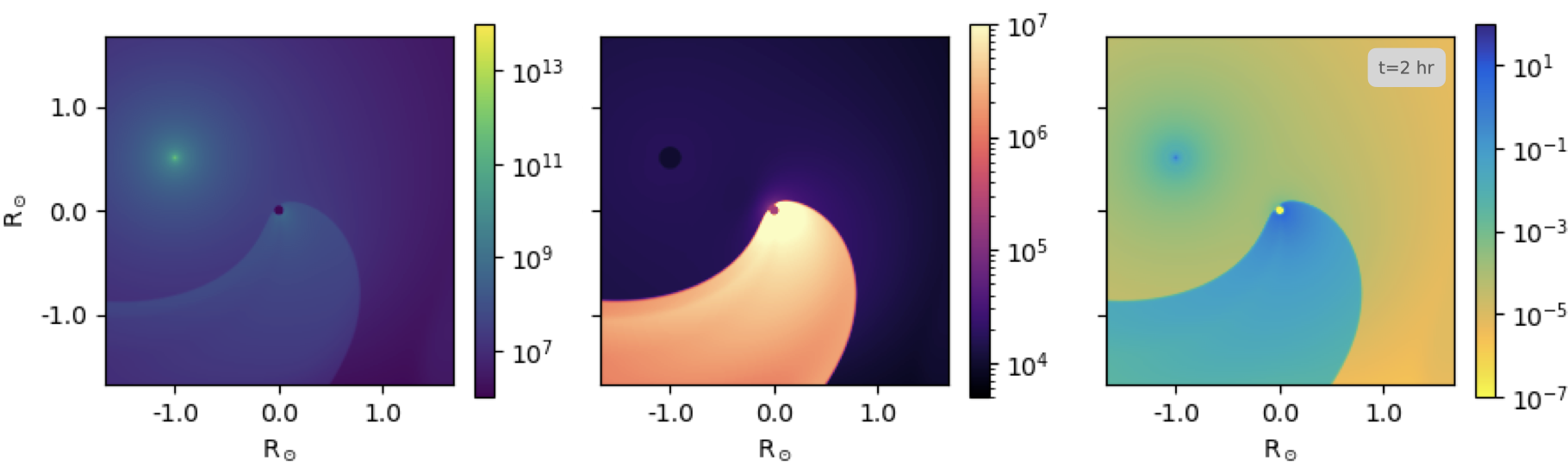}\\
\includegraphics[angle=0,width=\linewidth]{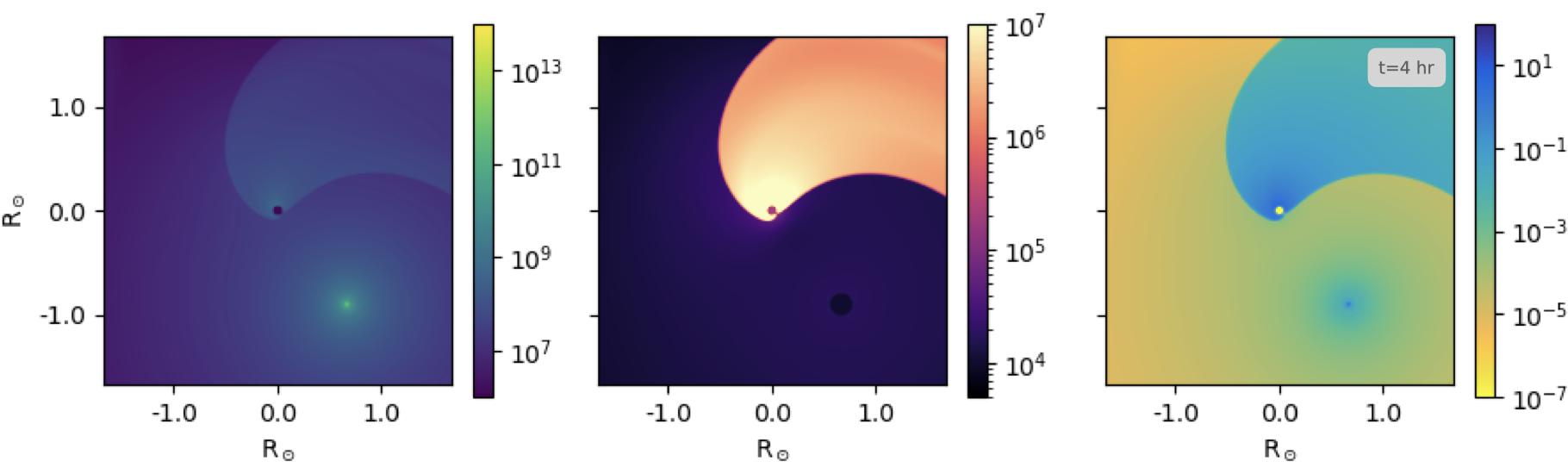}\\
\includegraphics[angle=0,width=\linewidth]{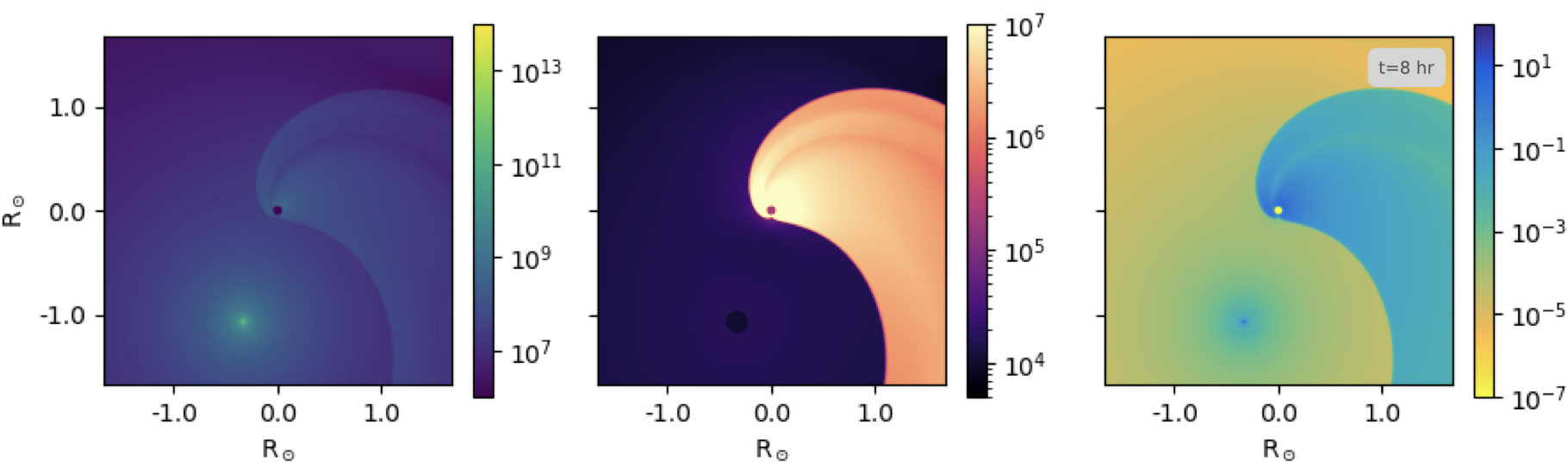}\\
\includegraphics[angle=0,width=\linewidth]{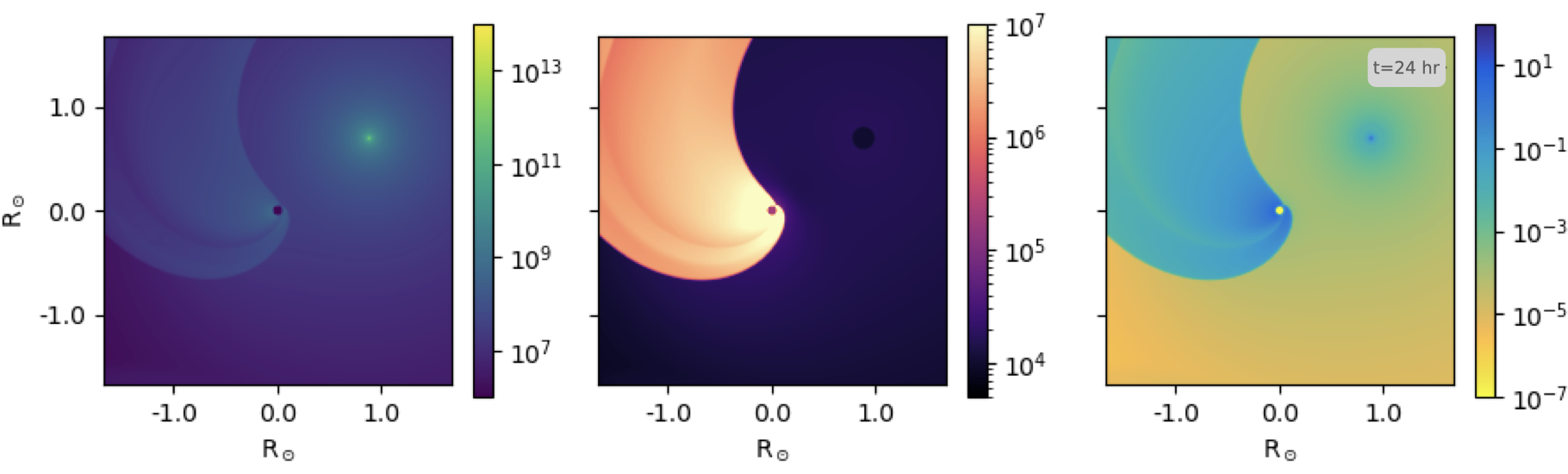}
\caption{Same as Fig.~\ref{fig:A_simfront} but for Run D.}
\label{fig:E_simfront}
\end{center}
\end{figure*}

\begin{figure*}
\begin{center}
\includegraphics[angle=0,width=\linewidth]{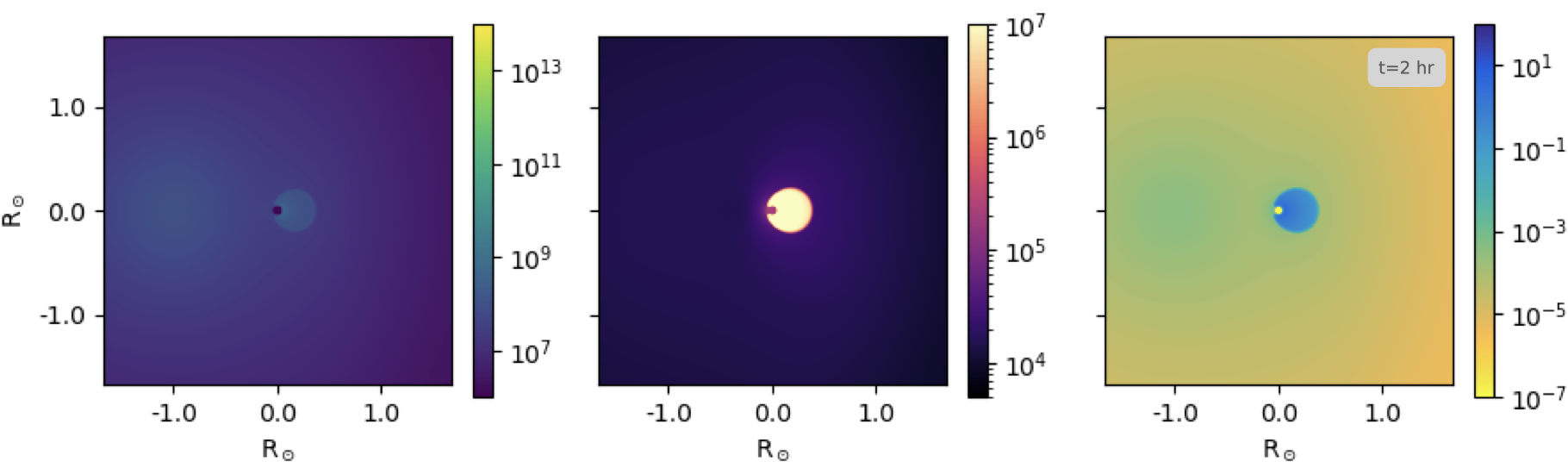}\\
\includegraphics[angle=0,width=\linewidth]{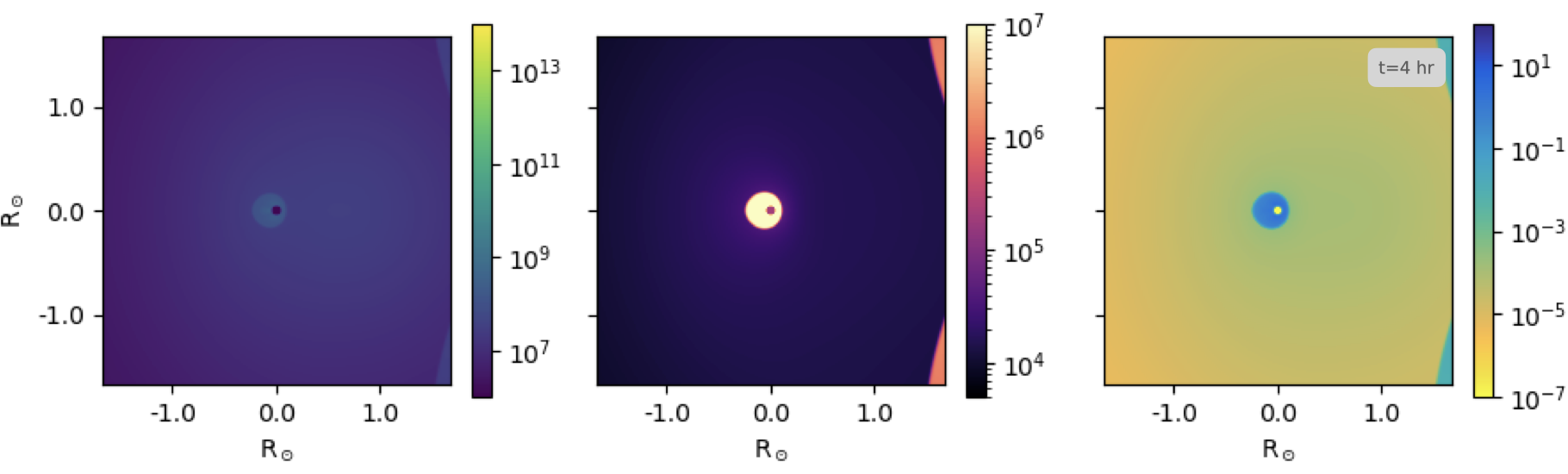}\\
\includegraphics[angle=0,width=\linewidth]{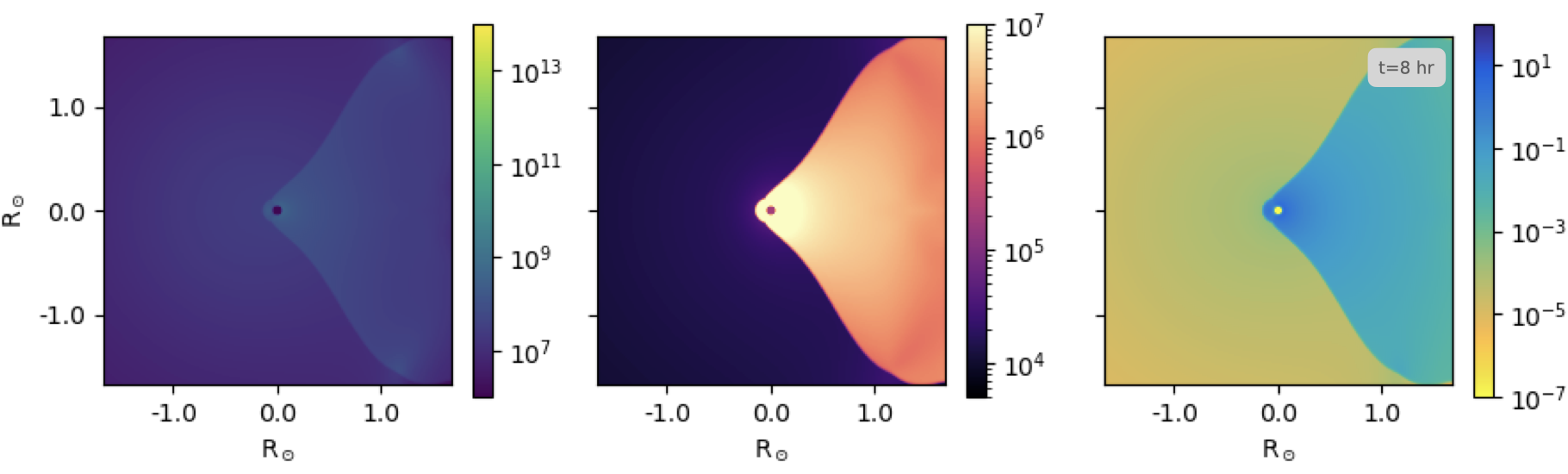}\\
\includegraphics[angle=0,width=\linewidth]{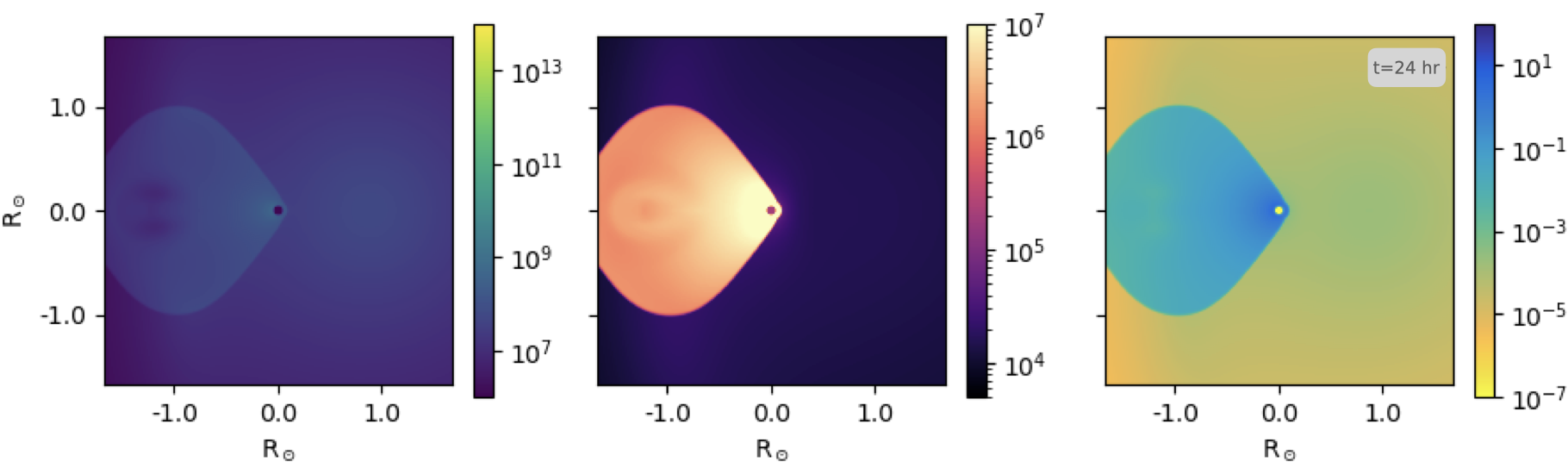}
\caption{Same as Fig.~\ref{fig:A_simside} but for Run D.}
\label{fig:E_simside}
\end{center}
\end{figure*}

\begin{figure*}
\begin{center}
\includegraphics[angle=0,width=0.45\linewidth]{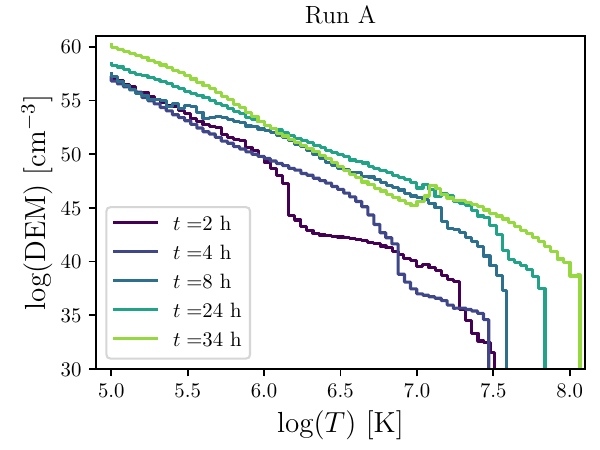}~
\includegraphics[angle=0,width=0.45\linewidth]{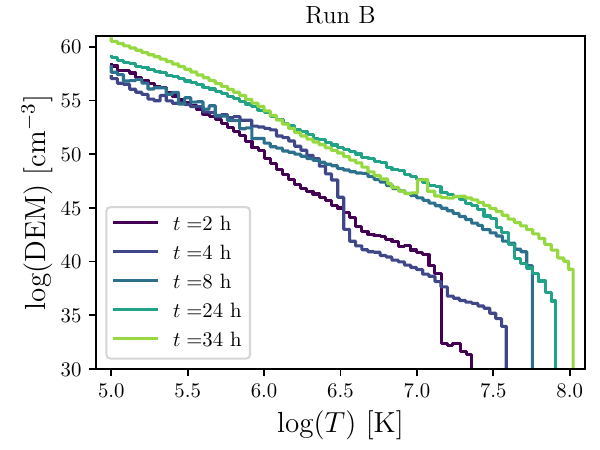}\\
\includegraphics[angle=0,width=0.45\linewidth]{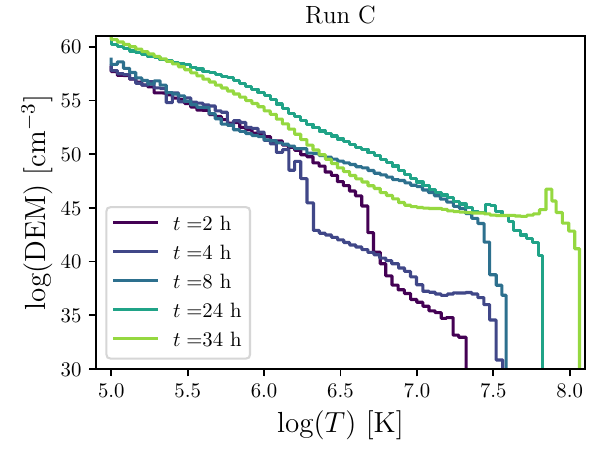}~
\includegraphics[angle=0,width=0.45\linewidth]{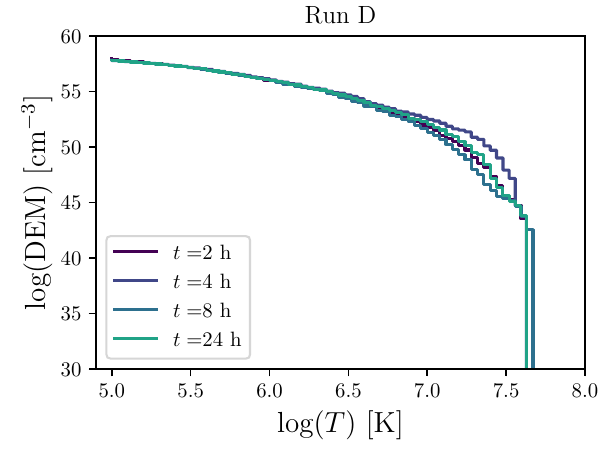}
\caption{Differential emission measure (DEM) versus gas temperature ($T$) obtained for Run A (top left), B (top right), C (bottom left) and D (bottom right). 
The different line colours correspond to simulation results for different evolution times from 2 to 34 hr.
}
\label{fig:DEM}
\end{center}
\end{figure*}

\begin{figure}
\begin{center}
\includegraphics[angle=0,width=0.94\linewidth]{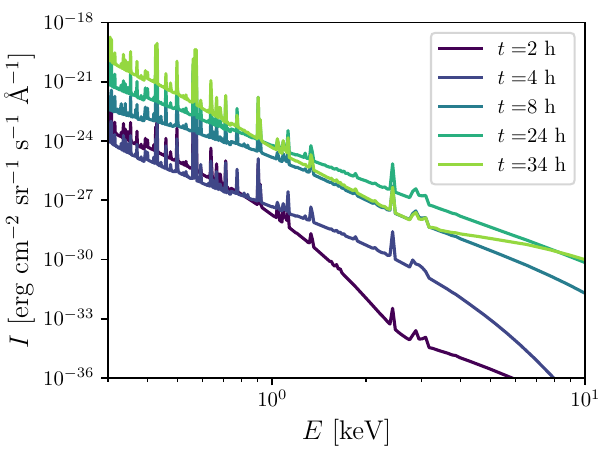}
\caption{Synthetic X-ray spectra extracted from Run~A using {\sc ChiantiPy}. Different colours show spectra obtained for different snapshots of the simulation.}
\label{fig:spectrum}
\end{center}
\end{figure}

The results from Runs~B and C, which have mass-loss rates 10 and 100 times larger than that adopted for Run~A, but similar wind velocities, are presented in Appendix~\ref{app:simulationsBC}. 
These simulations exhibit very similar morphological behaviour as that of Run~A and will not be described further here. 
The main difference is the disk mass, which is higher for simulations with higher mass-loss rate from the Jupiter-like planet. 
These differences will have impact on the production of X-ray-emitting material (see below).

Other simulations with higher outflow velocity from the Jupiter-like planet were also attempted. 
In particular, we show here a simulation with the same mass-loss rate as that of Run~A, but 10 times its outflowing velocity (600~km~s$^{-1}$), labelled as Run D. 
The numerical results are presented in Fig.~\ref{fig:E_simfront} and \ref{fig:E_simside} for the face-on and edge-on views, respectively. In Run D the fast wind produces a lobe-shaped shocked region with the WD at its head. 
This structure is similar to a bow-shock produced by a star moving through the ISM \citep[e.g.,][]{Green2022}, but deformed following the circular orbit of the Jupiter-like planet.

\subsection{Differential Emission Measure}

In order to compare our results with those obtained from X-ray observations, we need to generate synthetic X-ray observations from our numerical results. However, we first need to characterise the properties of the X-ray-emitting material in the simulations. A useful tool is the differential emission measure (DEM).

Following \citet{Toala2016,Toala2018}, we define the DEM for each time step of the simulation as
\begin{equation}
    \mathrm{DEM}(T_\mathrm{b}) =  \sum_{k, T_\mathrm{k} \in T_\mathrm{b}} n_\mathrm{e}^2 \Delta V_{k},
\end{equation}
\noindent where $n_\mathrm{e}$ is the electron number density in cell $k$, $\Delta V_k$ is the volume of cell $k$ and the sum is performed over cells with gas temperature falling in the bin whose central temperature is $T_\mathrm{b}$. In this work, we use logarithmic binning in the temperature range log($T$) = 5 to log($T$) = 9 in intervals of 0.04 dex (i.e., 100 bins). Lower temperatures are not taken into account in calculating the DEM, since their contribution to X-ray emission is negligible.

The DEM evolution with time corresponding to Runs~A, B, C and D is illustrated in Fig~\ref{fig:DEM}. As expected, the DEM evolution of Runs A, B and C are very similar, but models with a higher mass-loss rate have higher DEM values. 
In the case of Run~D, where no accretion disk is formed, the DEM profile does not change dramatically between different snapshots of the simulation.

\begin{figure*}
\begin{center}
\includegraphics[angle=0,width=0.5\linewidth]{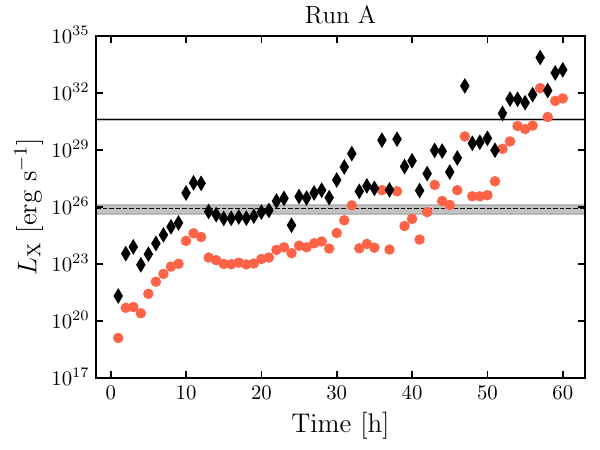}~
\includegraphics[angle=0,width=0.5\linewidth]{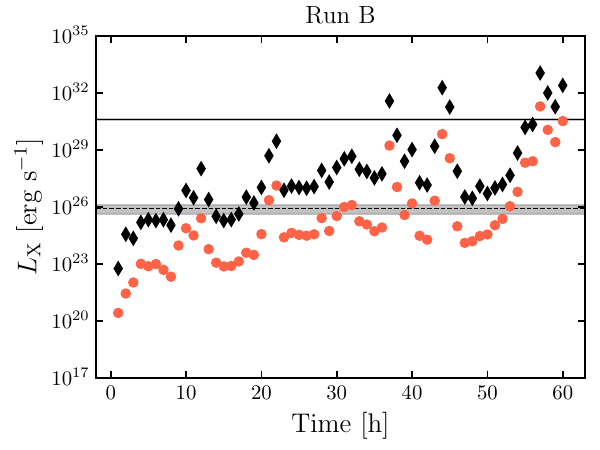}\\
\includegraphics[angle=0,width=0.5\linewidth]{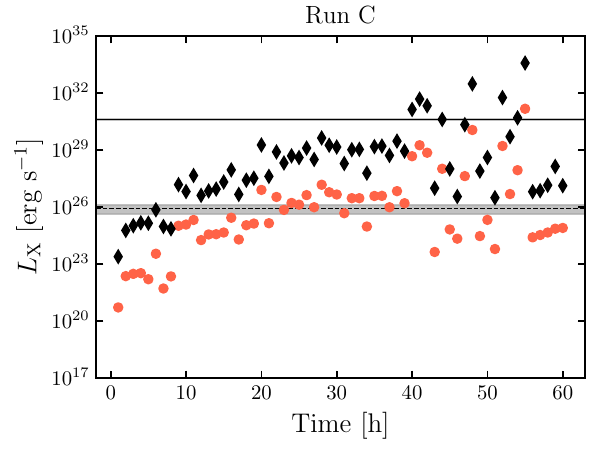}~
\includegraphics[angle=0,width=0.5\linewidth]{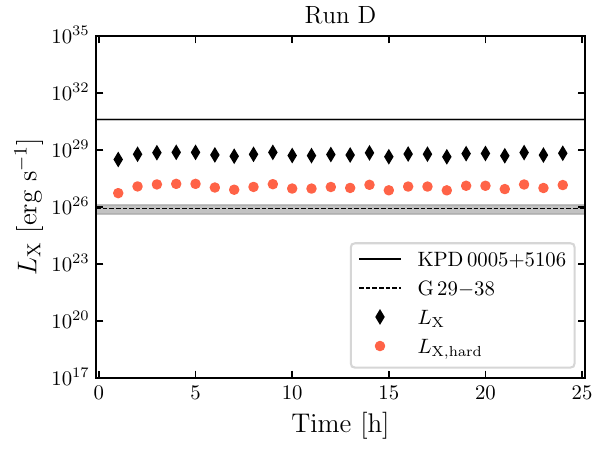}
\caption{Synthetic X-ray luminosity versus time (light curves) computed for Run A (top left), B (top right), C (bottom left) and D (bottom right). 
The (black) diamonds and (orange) dots show the results for the total ($L_\mathrm{X}$) and hard ($L_\mathrm{X,hard}$) X-ray luminosities. The horizontal solid- and dashed-line show the X-ray luminosities of KPD\,0005+5106 and G\,29-38, respectively.}
\label{fig:light_curves}
\end{center}
\end{figure*}

\subsection{Synthetic X-ray emission}

X-ray synthetic spectra are computed for each time and each simulation using the widely-tested {\sc Chianti} database through the {\sc ChiantiPy}\footnote{See \url{https://chiantipy.readthedocs.io/en/latest/}} routines \citep[version 14.0;][]{Dere2013,Dere2019}.

The calculation performed by {\sc Chianti} includes the contribution due to lines and to free-free, free-bound and two-photon continua. The abundances used to compute the spectra are those adopted for a standard ISM composition
in {\sc Cloudy} \citep[version 17.0;][]{Ferland2017}. A spectrum is generated for each temperature bin and weighted by its appropriate DEM value. The synthetic spectrum is simply the sum of all these individual contributions. All synthetic spectra are computed in the 0.3--7.0 keV energy range, corresponding to the 1.7--40~\AA\, wavelength range. The calculations are performed with a spectral bin of 0.1~\AA\, and the spectral line full width half maximum adopted in this work is 0.01~\AA.

Fig.~\ref{fig:spectrum} shows examples of the integrated spectrum that correspond to different snapshots of Run~A taking into account times between 2 and 34~hr of evolution. The different spectral shapes are a direct reflection of the contribution of the DEM profiles. For example, the early DEM profile of Run~A ($t$=2~h in the top left panel of Fig.~\ref{fig:DEM}) has a deficit of gas with temperatures higher than log($T$)=6.2 and this produces a spectrum dominated by emission with energies below 2~keV in Fig.~\ref{fig:spectrum}. 
On the other hand, the DEM at 4~hr of Run~A exhibits an increase of emission around log($T$)$\approx$6.5, which is reflected in an increase of emission for energies $>$2~keV in the spectrum as compared to that of the 2~hr of evolution. 
Similar spectral evolution patterns are exhibited by Runs B and C and are thus not shown here. 
Since the DEM profiles of Run~D do not exhibit great differences, their resultant synthetic spectra are almost identical at all times.

The synthetic spectra can be used to estimate total ($L_\mathrm{X}$) and hard X-ray luminosities ($L_\mathrm{X,hard}$). 
These were obtained by integrating the synthetic spectra in the 0.3--7.0~keV and 0.6--7.0~keV energy ranges for $L_\mathrm{X}$ and $L_\mathrm{X,hard}$, respectively, which are the spectral ranges used in the analysis of the observations by \citet{Chu2021} and \citet{EstradaDorado2023}.
This procedure is also performed over all spectra obtained for the different snapshots of the simulations. Fig.~\ref{fig:light_curves} presents the evolution of the X-ray luminosity with time (or simply light curves) of the four simulations presented here. The (black) diamonds represent the evolution of $L_\mathrm{X}$ whilst the (orange) bullets represent $L_\mathrm{X,hard}$.

The first thing to notice is that the light curves do not reflect the variability of the orbital period of the Jupiter-like planet adopted in the simulation, which is 4.7~hr. 
Although certain variability is exhibited by the light curves of Runs A, B and C, these do not match the adopted period and, in fact, these are not consistent among them. 
The variability of the light curves is thus suggested to be stochastic caused by periods of enhanced accretion associated with instabilities of the disk as in dwarf novae \citep{Warner1995}.
For discussion, Fig.~\ref{fig:light_curves} also shows in horizontal solid- and dashed-line the X-ray luminosity estimations corresponding to KPD\,0005+5106 and G\,29-38, respectively \citep[see][]{EstradaDorado2023,Chu2021}. 
We note that the X-rays in runs~A, B and C scales progressively in time, but it remains constant in simulation~D where an accretion disk is not formed. 
Those simulations that generate an accretion disk reproduce reasonably the X-ray luminosity and spectral properties of G\,29$-$38, but in the case of KPD\,0005+5106, the luminosity is exceeded in a few moments of the late simulations.
Due to the behaviour of the luminosities, those points that barely exceed the KPD\,0005+5106 luminosity could be interpreted as flares in the accretion disk, which generate a drop in the emission after the peak.  
These are thus
stochastic and not representative of the observed X-ray variable luminosity of KPD\,0005+5106.

\section{Discussion}
\label{sec:discussion}

The simulations presented here demonstrate that a WD is able to accrete material from a mass-losing planet and produce detectable X-ray emission. 
The synthetic X-ray fluxes predicted by our simulations are similar to those reported in the literature for X-ray-emitting evolved WDs accreting planetary material. Some details are worth discussing.

Runs A, B and C predict the formation of a relatively cold accretion disk. The X-ray-emitting material seems to be shock-heated at the inner regions of the disk and part of this material is subsequently ejected in the form of bipolar not-fully collimated outflow. 
The X-ray-emitting gas outflows and surrounds the accretion disk. In fact, the accretion disk has temperatures of a few times $10^{3}$~K, suggesting that this structure might contain hot dust.

It is interesting to note that the synthetic light curves presented in Fig.~\ref{fig:light_curves} do not reflect the orbital motion of the Jupiter-like planet around the WD adopted in the simulations. 
There is no clear pattern in the time evolution of the X-ray luminosity from Runs A, B or C. 
In fact, once the accretion disk is formed, the variability in the luminosity appears to be stochastic.

\begin{figure}
\begin{center}
\includegraphics[angle=0,width=\linewidth]{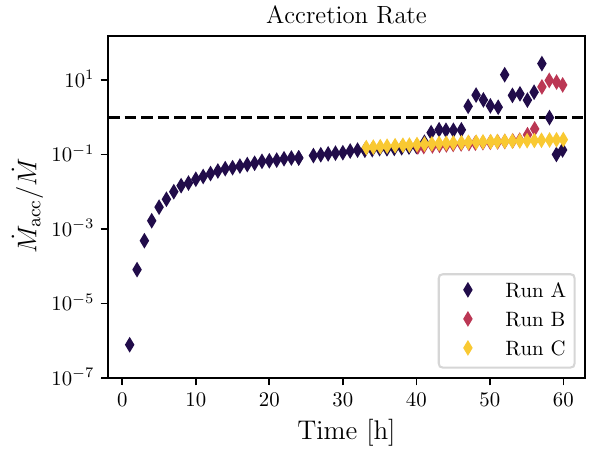}
\caption{Evolution of the normalized accretion rate  ($\dot{M}_\mathrm{acc}$) from the different simulations. Each simulation is normalized to its own mass-loss rate. The horizontal dashed-line represents an efficiency of 100 per cent.}
\label{fig:accretion}
\end{center}
\end{figure}

In order to assess the efficiency of the accretion process in our simulations, we calculated the accreted mass per unit of time of the simulation. This is illustrated for Runs A, B and C in Fig.~\ref{fig:accretion}. The evolution of the accreted mass per time of each simulation set is normalised to the planet's mass-loss rate. Fig.~\ref{fig:accretion} shows that at earlier times most of the mass is lost and not accreted by the WD given the isotropic nature of the wind from the planet. Once the accretion disk is formed ($\sim$10~hr) the efficiency reaches about 3 per cent. In the last 20~hr of the simulations, the accretion process has efficiency values between 18 and 25 per cent of the total mass lost by the planet. In later times of Runs A and B there is a couple of peaks in efficiency which we attribute to sudden accretion of material previously stripped from the planet. This figure confirms that the peaks 
in the light curves are also not related to the efficiency of the accretion process.  

The numerical results presented here are revealing and have consequences for KPD\,0005+5106. 
Our simulations suggest that a scenario in which the variable hard X-ray emission is produced by accretion from planetary material onto a WD might not be capable of explaining the observed properties of KPD\,0005+5106. 
Although at some times Runs A, B and C produce similar levels of X-ray emission, the presented simulations do not exhibit a clear periodic variability, which could be associated with that reported by the analysis of {\it Chandra} and {\it XMM-Newton} observations of this hot WD. 
Alternatively, the variable X-ray emission of KPD\,0005+5106 can imply the presence of a hot-spot in the surface of the rotating star.  
According to \citet{Dupuis2000} a hot-spot in the WD is possible if the surface of the star has a concentration of metals which generate a chemical inhomogeneity. 
The presence of a magnetic field channelling the accretion onto the magnetic poles can also produce hot-spots as in polar or intermediate-polar binary systems.   
The variability would then be generated by the rotation of the WD in combination with the presence of a hot spot. 
This would mean that the rotation period of KPD\,0005+5106 could be that reported for the hard X-ray emission (4.7~hr) which is consistent with the periods found for isolated WDs \citep[see, e.g.,][]{Hermes2017}.

The simulations presented here can also be used to discuss the X-ray emission detected from G\,29-38. The dashed line presented in Fig.~\ref{fig:light_curves} at 10$^{25}$~erg~s$^{-1}$ represents the X-ray luminosity from this WD \citep{EstradaDorado2023}. This means that the accretion disk scenario reproduces the emission of G\,29-38, which is a polluted WD \citep{KPS1997} with planetary debris \citep{Cunningham2022} falling into the WD. This WD is one of the first metal-polluted WD where IR excess was associated with an accretion disk \citep{Graham1990,Zuckerman1987}. The origin of such structure has been attributed to the presence of the disruption of an asteroid or a minor planet \citep{Jura2003,Reach2005}, but IR observations of this structure suggest that its chemical composition is better explained by material disrupted from a Jupiter-like planet according to \citet[][]{Reach2009}. In fact, a 4.4$\sigma$ detection of X-ray emission from G\,29-38 has been used to suggest that this WD is  accreting material from remnant planetary material \citep{Cunningham2022}. The latter has been corroborated by our team using {\it XMM-Newton} observations \citep[see][]{EstradaDorado2023}.
Except for Run~D, the other three simulations predict similar X-ray luminosities and the formation of a relatively cold accretion disk, similar to what is found in G\,29-38 (see Fig.~\ref{fig:light_curves}).

\section{Summary} 
\label{sec:summary}

We presented radiation-hydrodynamic simulations of the accretion of planetary material onto a WD. 
The simulations adopt the parameters (orbital period, accretion rate) estimated from the hard X-ray emission from the hot WD KPD\,0005+5106. 
The simulations reproduce the formation of a relatively cold accretion disk around a WD with the X-ray emitting material leaving this disk mostly along the polar directions. 
The simulations predict X-ray luminosity levels similar to those reported in the literature from observations of the metal-polluted DAZ WD G\,29$-$38, but marginally reproduce the hard X-ray emission for KPD\,0005+5106 for certain times.
The variability of the synthetic X-ray light curve does not follow the assumed orbital period, but is more likely stochastic. 
The X-ray emission basically grows with time until it reaches an asymptotic value when the accretion rate attains a steady value.

Consequently, we suggest that the variable hard X-ray emission detected from KPD\,0005$+$5106 has another origin. 
We propose that it might be due to the presence of a hot spot on its surface most likely associated with the stellar magnetic field. 
The X-ray variability period would then be associated with the rotation period, which is most likely also the orbital period of a close companion as a close binary system will establish synchronous rotation. 
Further optical and X-ray monitoring is urged in order to further assess these characteristics.

Finally, we note that although the presented simulations were not tailored to the specific case of G\,29-38, they are able to reproduce the observed X-ray luminosity level and the production of an accretion disk formed by planetary material as previously suggested. 
Given the low temperature of the accretion disk, we predict that it would contain host dust, which is another observable characteristic in G\,29-38.

%%%%%%%%%%%%%%%%%%
% Acknowledgements
%%%%%%%%%%%%%%%%%%
\section*{Acknowledgements}

We thank the anonymous referee for comments and suggestions that improved our original manuscript.
SE-D thanks Consejo Nacional de Ciencia y Tecnolog\'{i}a (CONACyT, Mexico) for a student grant and technical support provided by A. López-Maldonado. SE-D and JAT acknoledge funding from the UNAM DGAPA PAPIIT project IA101622. JAT thanks support from the Marcos Moshinksy Foundation and the Visiting-Incoming programme of the IAA-CSIC through the Centro de Excelencia Severo Ochoa (Spain). MAG acknowledges support of grant PGC 2018-102184-B-I00 of the Ministerio de Educaci\'{o}n, Innovaci\'{o}n y Universidades cofunded with FEDER funds. 
RFM aknowledges the funding from UNAM DGAPA postdoctoral fellowship.
YHC acknowledges the support of grants NSTC 111-2112-M-001-063- and NSTC 112-2112-M-001-065- from the National Science and Technology Center of Taiwan. 
VL acknowledges the support of CONAHCyT. VL and AE acknowledge support from PAPIIT-UNAM grant IN113522. In memoriam of Dr.\ Alejandro Raga, co-author of the {\sc guacho} code, who sadly passed away recently. 
This work has made extensive use of NASA's Astrophysics Data System (ADS).

\section*{Data availability}
The results of our simulations presented in this work are available in the article. Our results will be shared on reasonable request to the first author.

%%%%%%%%%%%%%%
% Bibliography
%%%%%%%%%%%%%%

\appendix
%\FloatBarrier

\section{Simulations B and C}
\label{app:simulationsBC}

\begin{figure*}
\begin{center}
\includegraphics[angle=0,width=0.9\linewidth]{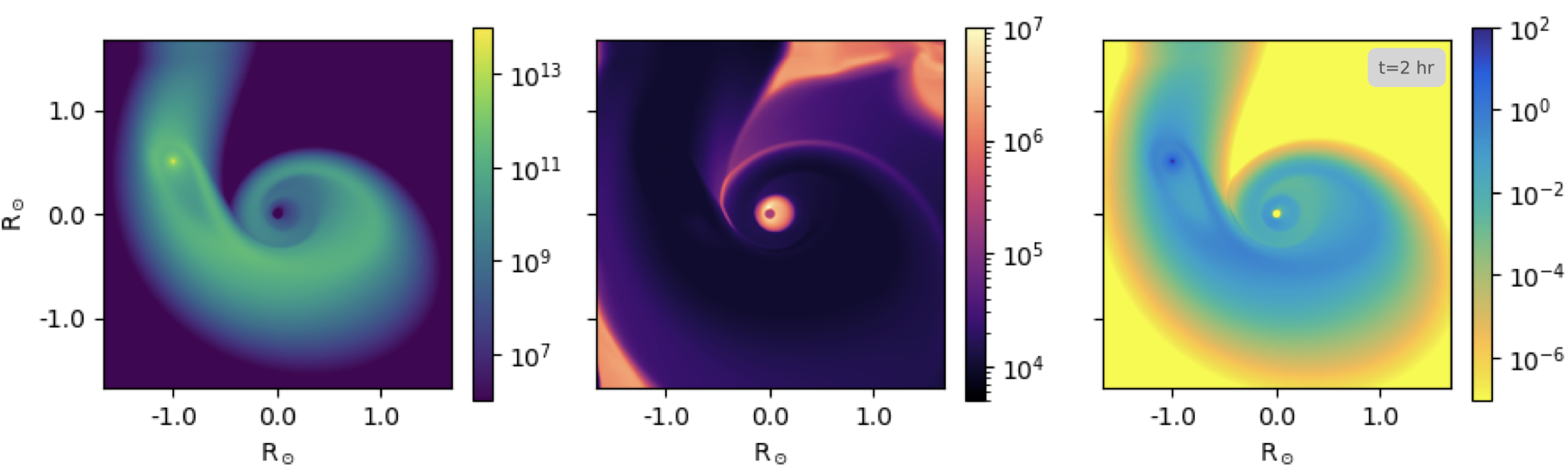}\\
\includegraphics[angle=0,width=0.9\linewidth]{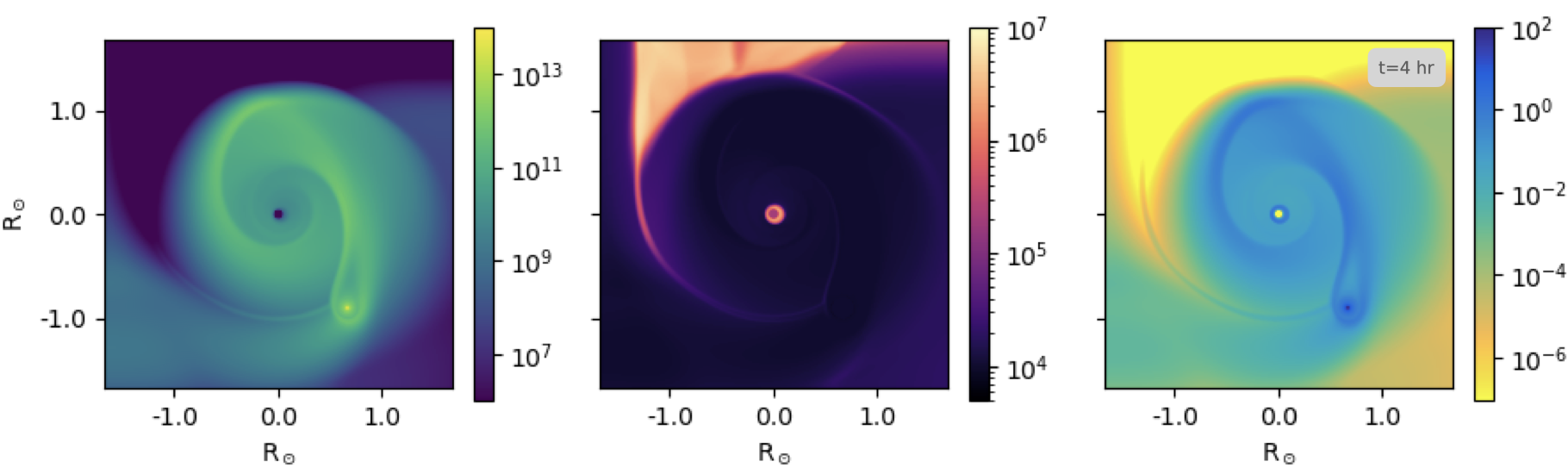}\\
\includegraphics[angle=0,width=0.9\linewidth]{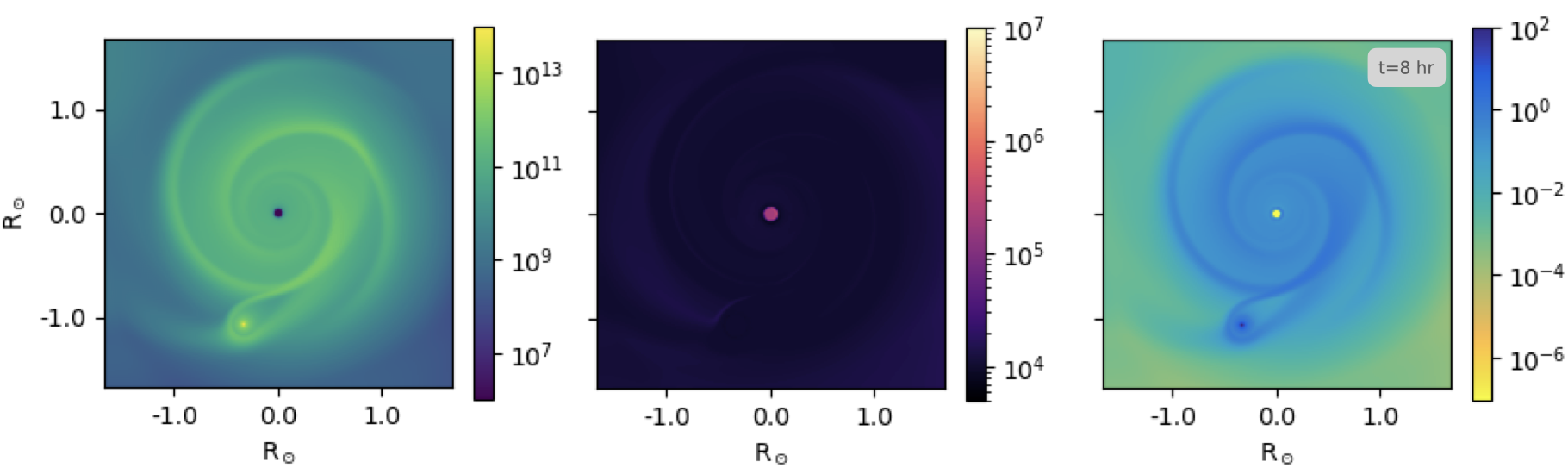}\\
\includegraphics[angle=0,width=0.9\linewidth]{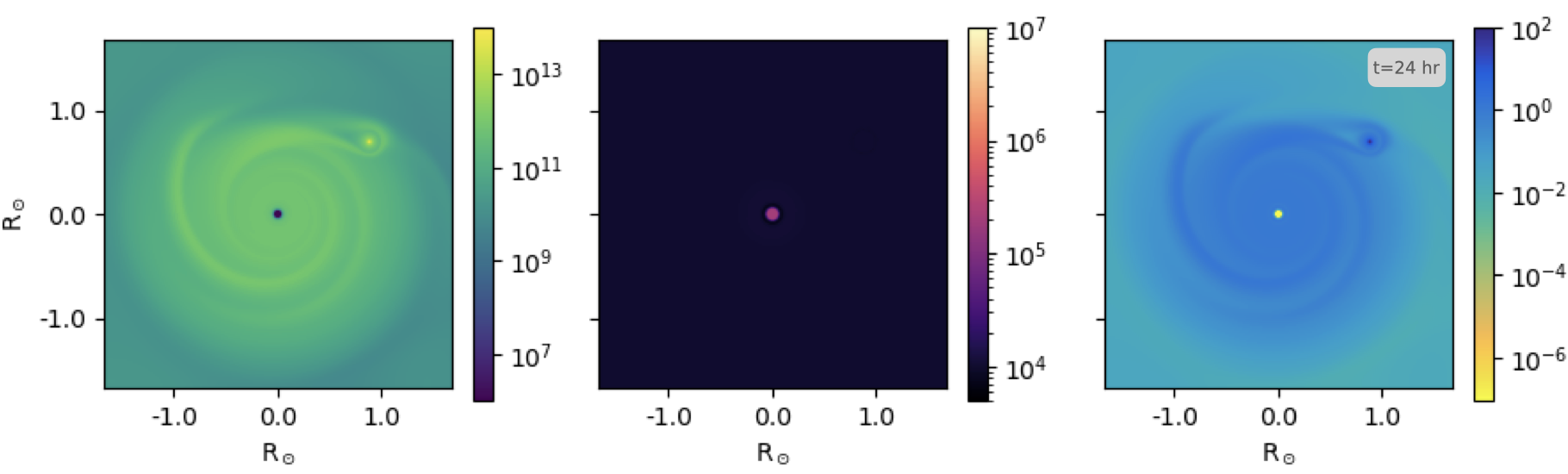}
\caption{Face-on view of the middle plane of the number density $n$ (cm$^{-3}$ - left), temperature $T$ (K - middle) and pressure (erg~cm$^{-3}$ - right) of Run B. From top to bottom we show show the results after 2, 4, 8 and 24 hrs of evolution.}
\label{fig:B_simfront}
\end{center}
\end{figure*}

\begin{figure*}
\begin{center}
\includegraphics[angle=0,width=0.9\linewidth]{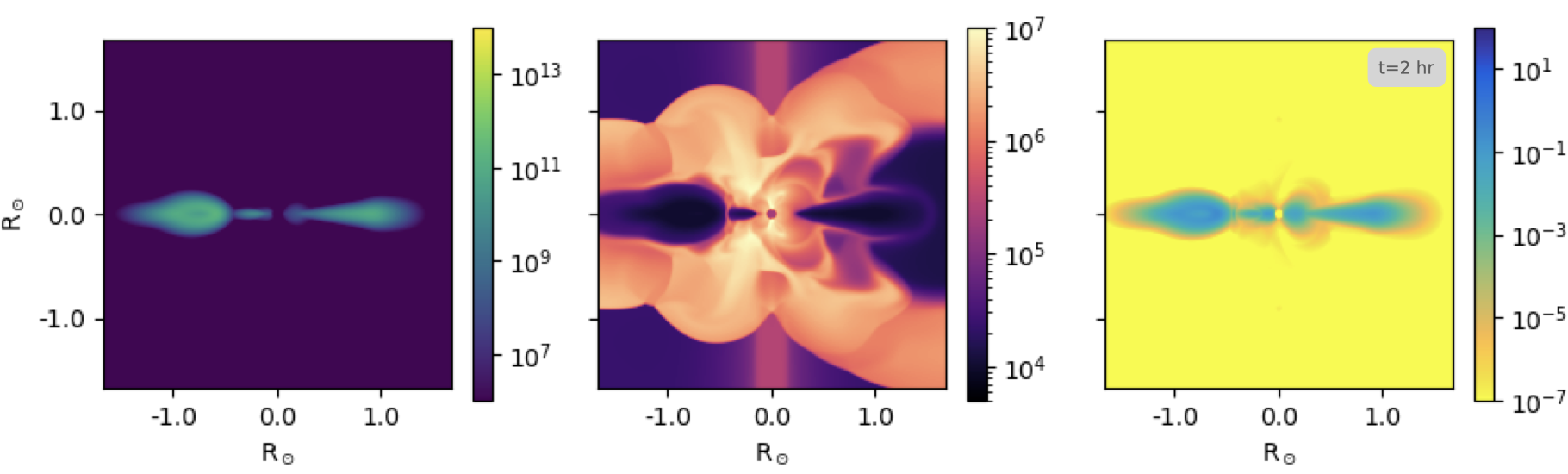}\\
\includegraphics[angle=0,width=0.9\linewidth]{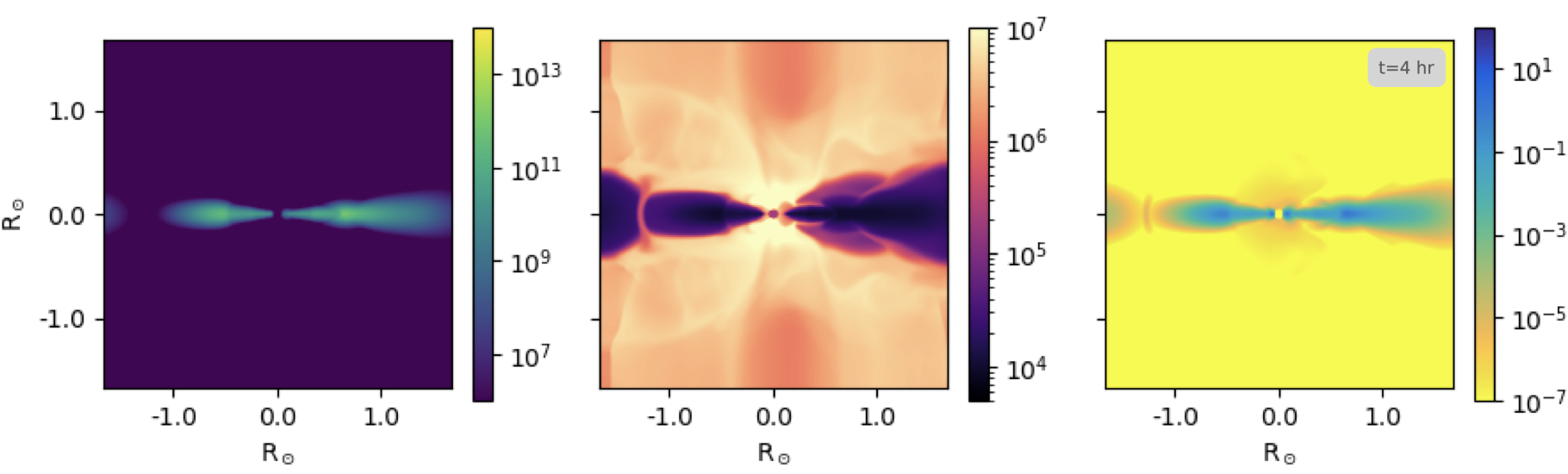}\\
\includegraphics[angle=0,width=0.9\linewidth]{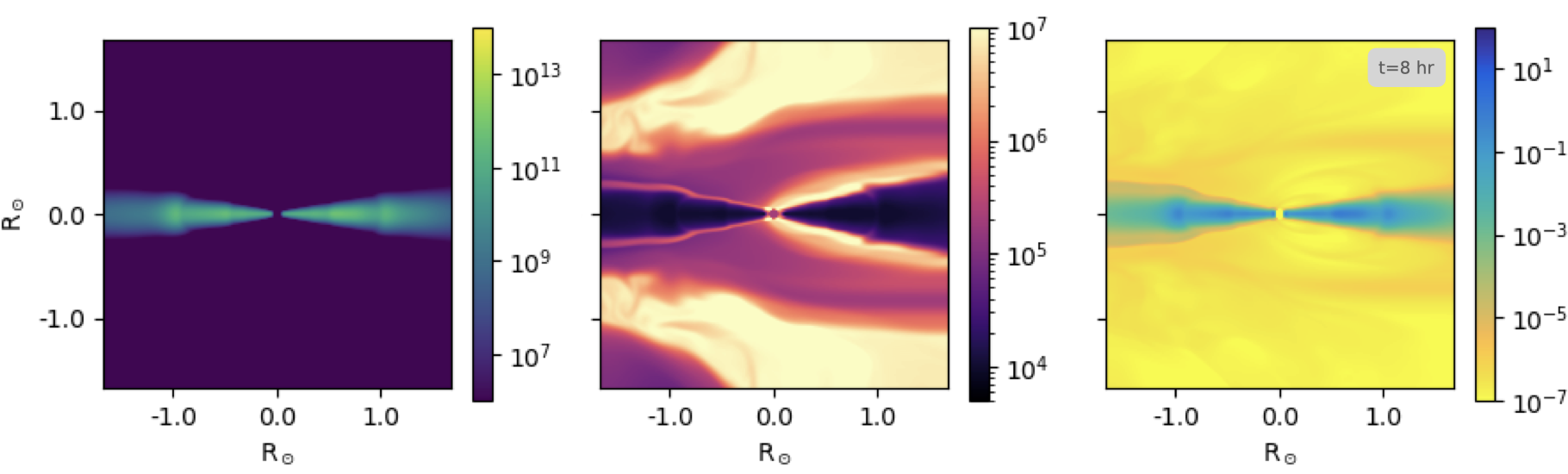}\\
\includegraphics[angle=0,width=0.9\linewidth]{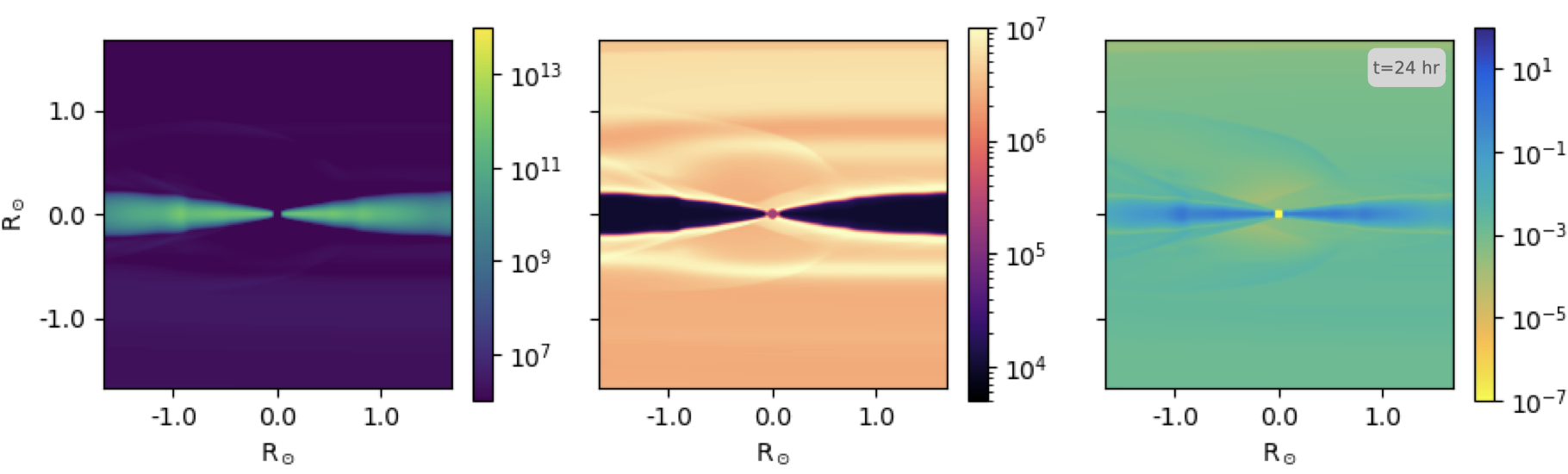}
\caption{Edge-on view of the middle plane of the number density $n$ (cm$^{-3}$ - left), temperature $T$ (K - middle) and pressure (erg~cm$^{-3}$ - right) of Run B. From top to bottom we show show the results after 2, 4, 8 and 24 hrs of evolution.}
\label{fig:B_simside}
\end{center}
\end{figure*}

\begin{figure*}
\begin{center}
\includegraphics[angle=0,width=0.9\linewidth]{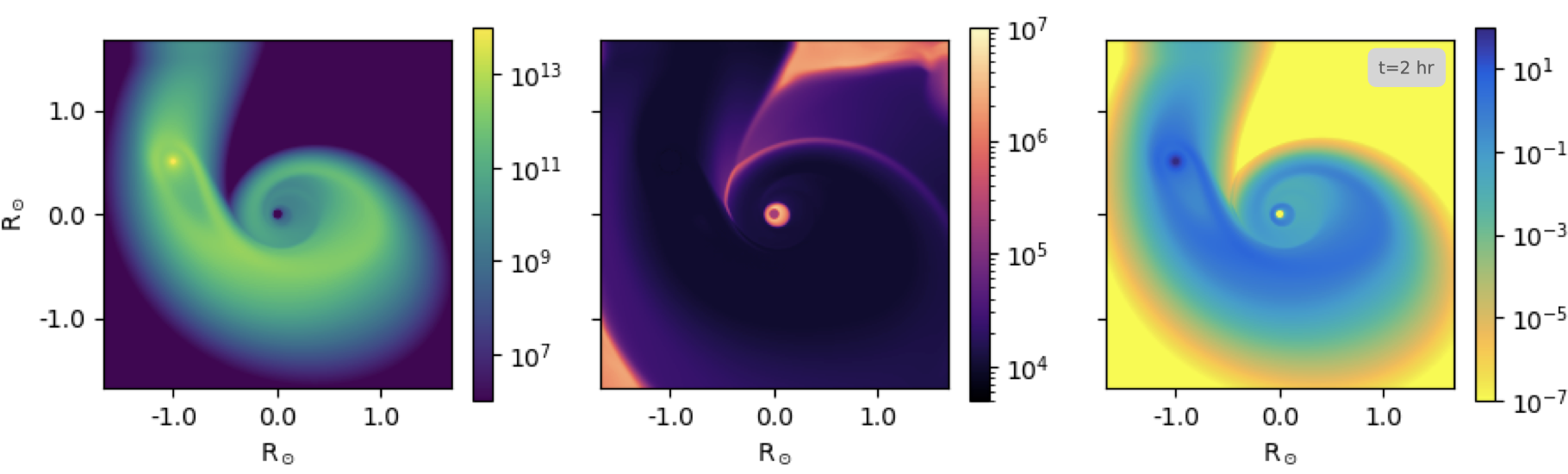}\\
\includegraphics[angle=0,width=0.9\linewidth]{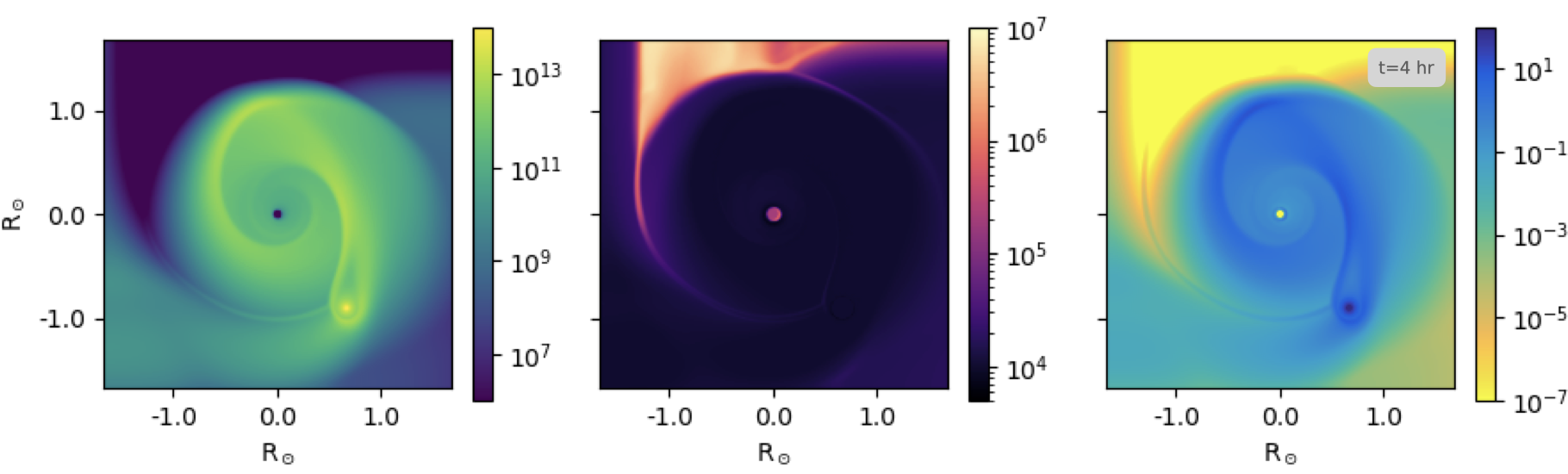}\\
\includegraphics[angle=0,width=0.9\linewidth]{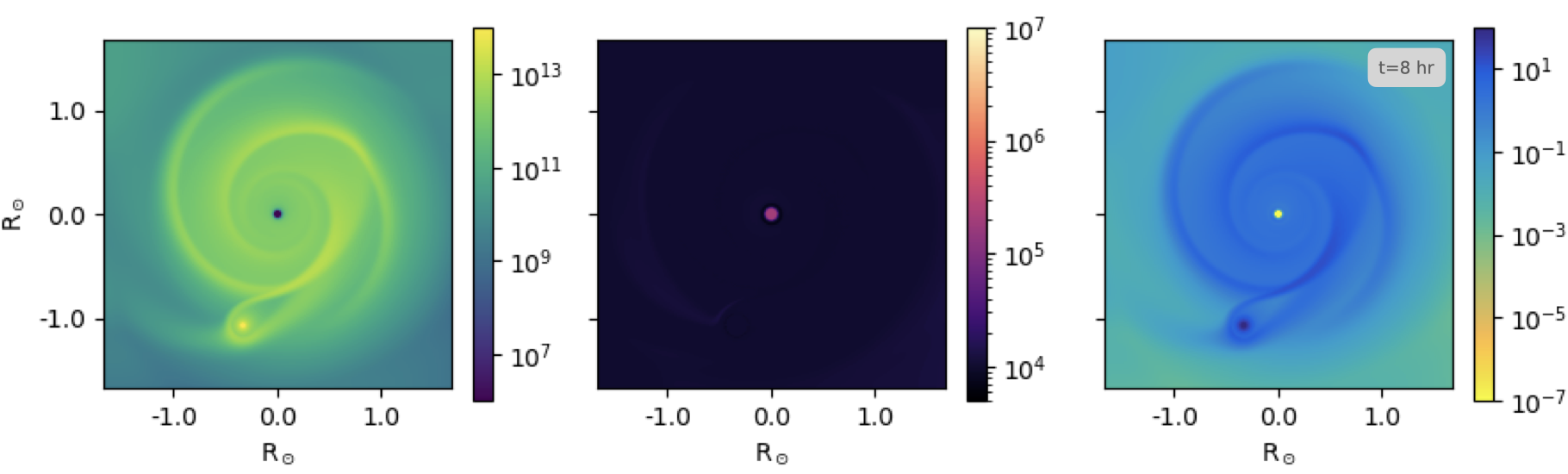}\\
\includegraphics[angle=0,width=0.9\linewidth]{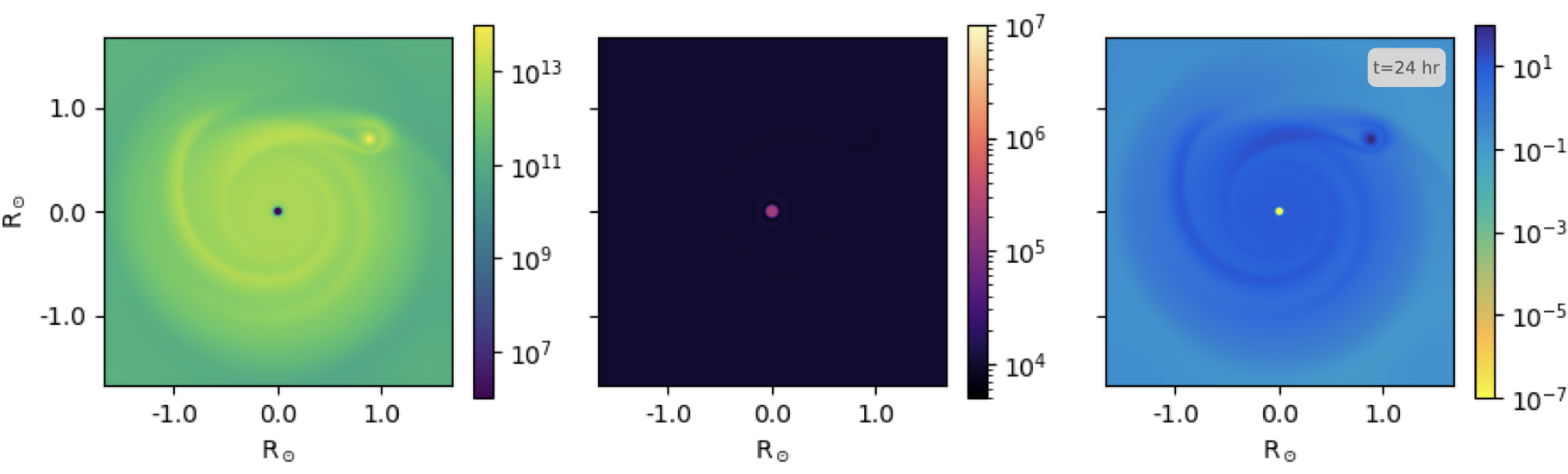}
\caption{Face-on view of the middle plane of the number density $n$ (cm$^{-3}$ - left), temperature $T$ (K - middle) and pressure (erg~cm$^{-3}$ - right) of Run C. From top to bottom we show show the results after 2, 4, 8 and 24 hrs of evolution.}
\label{fig:C_simfront}
\end{center}
\end{figure*}

\begin{figure*}
\begin{center}
\includegraphics[angle=0,width=0.9\linewidth]{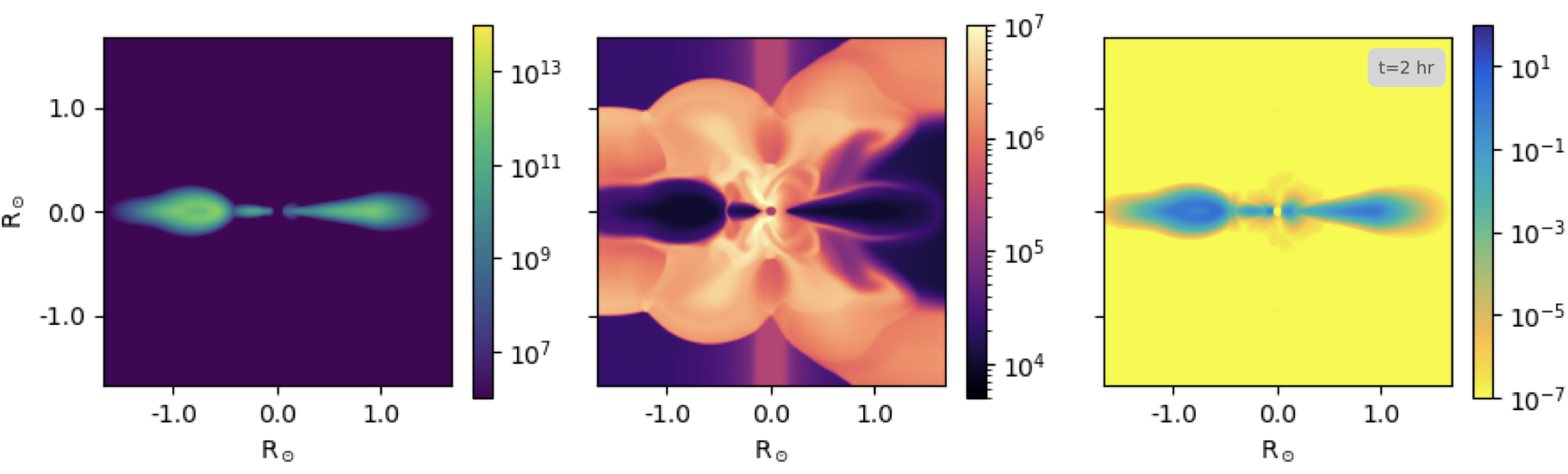}\\
\includegraphics[angle=0,width=0.9\linewidth]{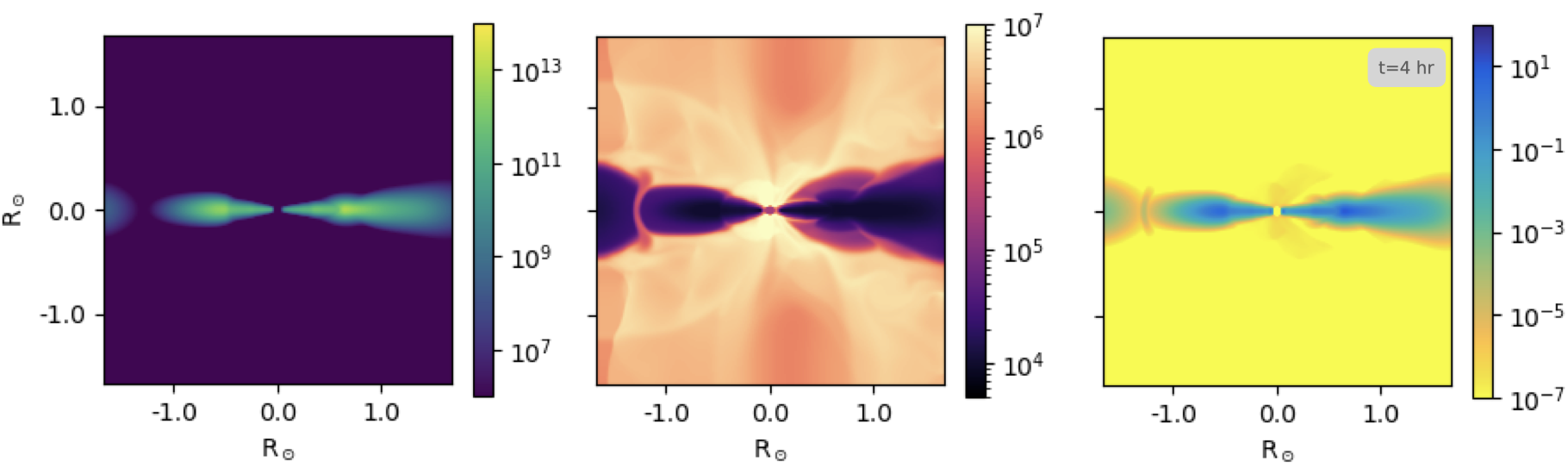}\\
\includegraphics[angle=0,width=0.9\linewidth]{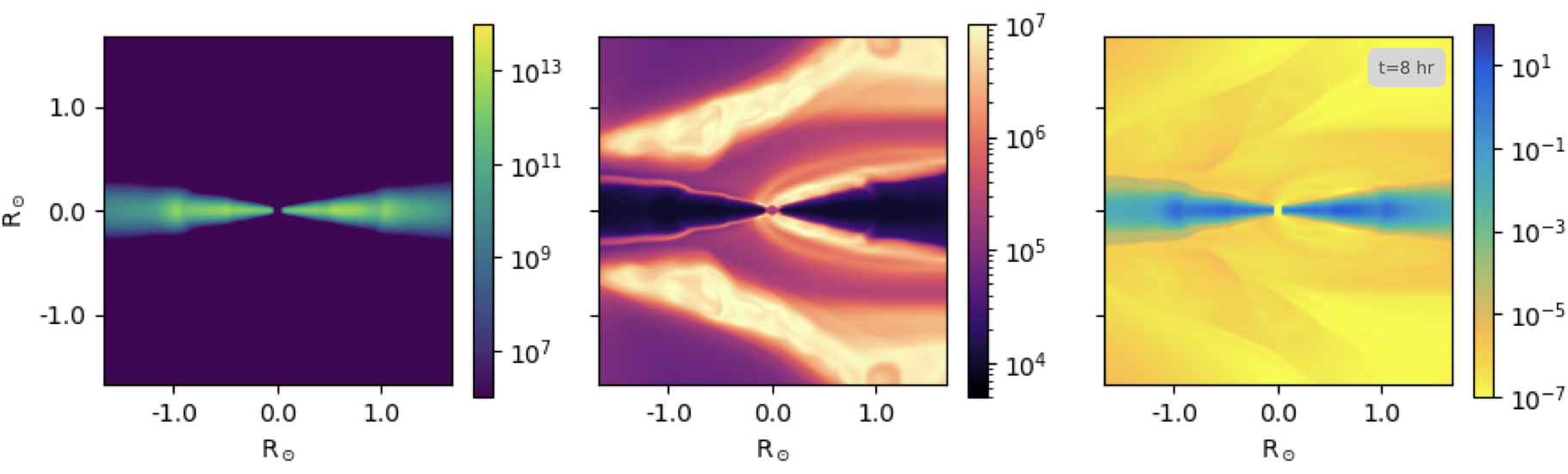}\\
\includegraphics[angle=0,width=0.9\linewidth]{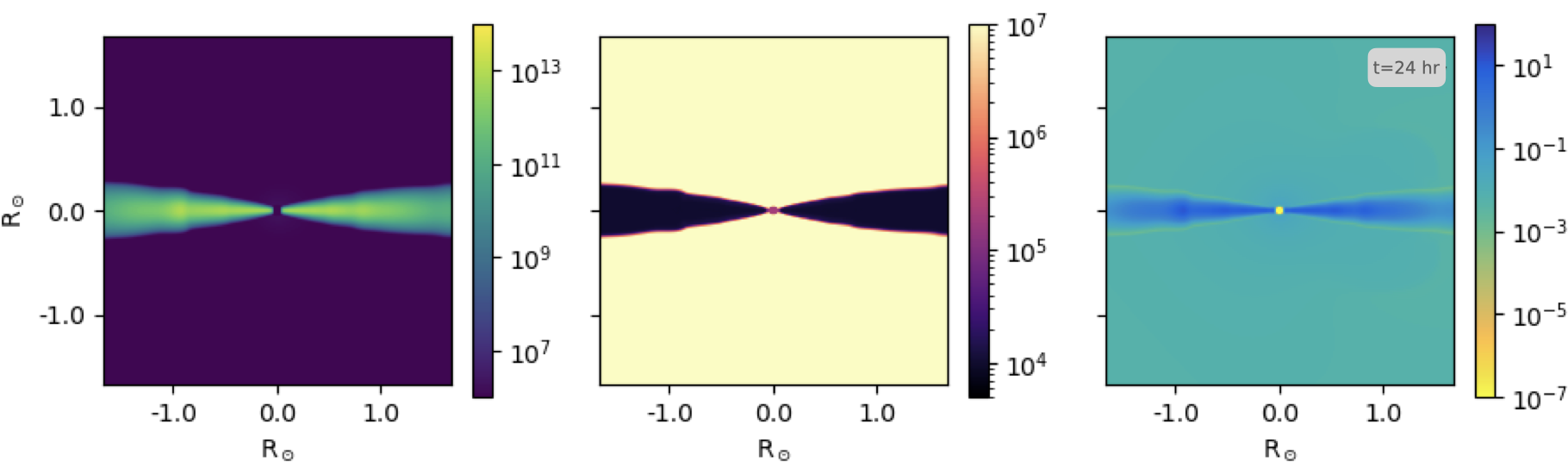}
\caption{Edge-on view of the middle plane of the number density $n$ (cm$^{-3}$ - left), temperature $T$ (K - middle) and pressure (erg~cm$^{-3}$ - right) of Run C. From top to bottom we show show the results after 2, 4, 8 and 24 hrs of evolution.}
\label{fig:C_simside}
\end{center}
\end{figure*}

In this appendix we present the hydrodynamical results of Runs B (Fig.~\ref{fig:B_simfront} and \ref{fig:B_simside}) and C (Fig.~\ref{fig:C_simfront} and \ref{fig:C_simside}). These simulations are very similar to those of Run~A presented in Section~\ref{sec:results} exhibiting the formation of the accretion disk around the WD (located at the centre of the numerical grid). Spiral arms of material spiraling in towards the accreting WD is disclosed by the face-on panels (see Fig.~\ref{fig:B_simfront} and \ref{fig:C_simfront}). Given that Runs B and C have 10 and 100 times higher mass-loss rate than Run~A, respectively, the density of the accretion disks are higher that in that model. No other remarkable differences are produced.

\end{document}